%% file: draft.tex
\documentclass[%
 reprint,
 amsmath,
 amssymb,
 aps,
 nofootinbib,
 preprintnumbers
]{revtex4-2}

\usepackage{physics}
\usepackage{graphicx}
\usepackage{centernot}       
\usepackage{dcolumn}
\usepackage{bm}
\usepackage[parfill]{parskip}
\usepackage{afterpage}
\usepackage[usenames,dvipsnames]{xcolor}
\usepackage{hyperref}
\hypersetup{
	colorlinks=true,
	urlcolor=Maroon,
	linkcolor=RoyalBlue,
	citecolor=Maroon,
	pdftitle={},
	pdfauthor={},
	pdfdisplaydoctitle=true,
	pdfstartview=FitH,
	linktocpage=true
}

\usepackage{amsmath}
\usepackage{dsfont}
\usepackage{amsfonts}
\usepackage{amssymb}
\usepackage{mathtools}
\usepackage{microtype}
\usepackage{bm}
\usepackage{tikz, pict2e}
\usepackage{lmodern}
\usepackage{tkz-euclide}
\usepackage[utf8]{inputenc}
\usepackage{blkarray}
\usepackage{bigstrut}
\usepackage{nccmath}
\usepackage{enumitem}
\usepackage{float}
\usepackage{algorithm}
\usepackage{braket}
\usepackage{dsfont}
\usepackage[export]{adjustbox}
\usepackage[english]{babel}
\usepackage{stmaryrd}
\usepackage{orcidlink}
\usepackage{tikz}
\usepackage{bbm}
\usepackage{array}
\usepackage{comment} 
\usepackage{orcidlink}
\usetikzlibrary{calc,backgrounds, intersections}
\usetikzlibrary{arrows,shapes,positioning,shadows,trees}
\usetikzlibrary{mindmap}
\usetikzlibrary{decorations.pathreplacing, calligraphy}
\usetikzlibrary{decorations.pathmorphing}
\usetikzlibrary{decorations.markings}
\usetikzlibrary{matrix}

\tikzstyle arrowstyle=[scale=1]
\tikzstyle directed=[postaction={decorate,decoration={markings,
    mark=at position .5 with {\arrow[arrowstyle]{stealth}}}}]
\tikzstyle reverse directed=[postaction={decorate,decoration={markings,
    mark=at position .5 with {\arrowreversed[arrowstyle]{stealth};}}}]

\definecolor{redb}{rgb}{0.700, 0.000, 0.300}
\DeclareMathAlphabet\mathbfcal{OMS}{cmsy}{b}{n}

\usepackage{nicematrix}
\usetikzlibrary{calc}
\usetikzlibrary{patterns}
\usetikzlibrary{shapes.misc}
\usetikzlibrary{
    decorations.pathmorphing,
    decorations.pathreplacing,
    decorations.shapes,
    decorations.markings,
}
\usepackage{fancyhdr}

\makeatletter
\newlength{\apb@width}
\newcommand{\autoparbox}[2][c]{\settowidth{\apb@width}{#2}\parbox[#1]{\apb@width}{#2}}

\makeatother

\definecolor{green1}{HTML}{244819}
\definecolor{cyan1}{HTML}{37cdaa}
\definecolor{blue1}{HTML}{5d7ac4}
\definecolor{red1}{HTML}{921818}
\definecolor{purple1}{HTML}{53047A}
\definecolor{orange1}{HTML}{e07229}
\definecolor{yellow1}{HTML}{edcb52}
\definecolor{gr}{gray}{0.7}
\definecolor{gr1}{gray}{0.7}

\usepackage{pgf}

\newcommand{\oldornew}[1]{}

\newcommand{\bk}{\kappa}
\newcommand{\gm}{\chi}
\newcommand{\q}{q}
\DeclareMathAlphabet\mathbfcal{OMS}{cmsy}{b}{n} 

\newcommand\scalemath[2]{\scalebox{#1}{\mbox{\ensuremath{\displaystyle #2}}}}

\begin{document}

\preprint{MPP-2024-174}

\title{On one-loop corrections to the Bunch-Davies wavefunction of the universe}

\author{Paolo Benincasa$\,^{a,b,\orcidlink{0000-0002-6717-7922}}$}
\email{pablowellinhouse@anche.no}
\author{Giacomo Brunello$\,^{c,d,e,\orcidlink{0009-0004-4788-738X}}$}
\email{giacomo.brunello@phd.unipd.it}
\author{Manoj K. Mandal$\,^{c,d,\orcidlink{0000-0003-0850-7685}}$}
\email{manojkumar.mandal@pd.infn.it}
\author{Pierpaolo Mastrolia$\,^{c,d,\orcidlink{0000-0001-9711-7798}}$}
\email{pierpaolo.mastrolia@unipd.it}
\author{Francisco Vaz\~ao$\,^{a\orcidlink{0000-0002-6518-9786}}$}
\email{fvvazao@mpp.mpg.de}

\affiliation{$^a$ Max-Planck-Institut  f{\"u}r Physik, 
     Werner-Heisenberg-Institut, D-80805 München, Germany}
\affiliation{$^b$Instituto Galego de F\'{i}sica de Altas Enerx\'{i}as IGFAE, Universidade de Santiago de Compostela, E-15782 Galicia-Spain}
\affiliation{$^c$Dipartimento di Fisica e Astronomia, Universita di Padova, Via Marzolo 8, 35131 Padova, Italy}
\affiliation{$^d$INFN, Sezione di Padova,
Via Marzolo 8, I-35131 Padova, Italy.}
\affiliation{$^e$Institut de Physique Théorique, CEA, CNRS, Université Paris-Saclay, F–91191 Gif-sur-Yvette cedex, France}

\begin{abstract}
Understanding the loop corrections to cosmological observables is of paramount importance for having control on the quantum consistency of a theory in an expanding universe as well as for phenomenological reasons. In the present work, we begin with a systematic study of such corrections in the context scalar toy models whose perturbative Bunch-Davies wavefunction enjoys an intrinsic definition in terms of {\it cosmological polytopes}, 
focusing on one-loop graphs. 
Owing to the underlying twisted period integral representation they admit, 
their combinatorial structure 
along with their vector space structure, emerging from polynomial ideals algebra and intersection theory, 
are exploited to set-up and analyse the differential equations that the two- and three-site one-loop corrections have to satisfy upon variation of the external kinematic variables. We find that, while the two-site contribution can be written in terms of multiple-polylogarithms, this is no longer true for the three-site case, for which elliptic structures appear. As a non-trivial check, we consider the scattering amplitude limit, recovering the known result in terms of polylogarithms only.\\ 
\end{abstract}

\maketitle

\allowdisplaybreaks
\raggedbottom
\input{sections/introduction}
\input{sections/cosmologicalintegrals}
\input{sections/method}

\input{sections/Fullintegration}
\input{sections/PartialFractioning}
\input{sections/applications}
\input{sections/outlooks}


\acknowledgments

It is pleasure to thank Vsevolod Chestnov, Leonardo De La Cruz, Jungwon Lim, Subodh Patil, Guilherme Pimentel, Andrzej Pokraka, Prashanth Raman, Tom Westerdijk for insightful discussions. G.B. thanks John Joseph Carrasco and the Northwestern University for hospitality during the completion of this work. P.M. acknowledges interesting discussions emerging at the workshop {\it Scattering Amplitudes and Cosmology} at ICTP Trieste, April 2023, on the relations between the calculus of cosmological integrals and intersection theory based methods here presented. G.B. and F.V wish to acknowledge interesting discussions emerging at the {\it 6th School of Analytic Computing in High-Energy and Gravitational Theoretical Physics} at Atrani, October 2023, which triggered the development of this project.
P.B. would like to thank Dieter Lust and Gia Dvali for making contributing to this work possible as well as the Instituto Galego de F{\'i}sica de Altas Enerx{\'i}as of the Universidade de Santiago de Compostela for hospitality during the whole development of this work. . P.B. would like to thank the University of Padova for hospitality while this work was in progress. G.B. research is supported by the European Research Council, under grant ERC–AdG–88541. G.B. research is supported by Università Italo-Francese, under grant Vinci. F.V. research is funded by the European Union (ERC, UNIVERSE PLUS, 101118787). Views and opinions expressed are however those of the author(s) only and do not necessarily reflect those of the European Union or the European Research Council Executive Agency. Neither the European Union nor the granting authority can be held responsible for them.

\appendix
\input{sections/appendix1}
\input{sections/appendix2}

\bibliographystyle{apsrev4-1}
\bibliography{references}

\end{document}

%% file: sections/introduction.tex
\section{\label{sec:intro}Introduction}

Correlation functions defined at the space-like boundary of a quasi-de Sitter space-time are of crucial importance in cosmology, since, on the one hand, they encode the physics of the inflationary era, and, on the other one, they set the initial conditions for the evolution that from the reheating leads to the structures we can observe today in our universe. How to extract fundamental physics out of them represents still a puzzle that has not been completely solved, despite the recent progress in understanding the mathematical structure as well as developing more insightful computational techniques for both these quantities and the Bunch-Davies wavefunction associated to the probability distribution underlying them -- see \cite{Baumann:2022jpr, Benincasa:2022gtd} and references therein, as well as \cite{Pinol:2023oux, Benincasa:2024leu}.

Such advances mostly concern perturbation theory \footnote{At non-perturbative level, some progress has been made mostly in the strict de Sitter set-up \cite{Hogervorst:2021uvp, DiPietro:2021sjt, Penedones:2023uqc, Loparco:2023rug, DiPietro:2023inn}.}, with explicit results obtained mainly at tree-level (for recent works see for example \cite{Anninos:2014lwa, Arkani-Hamed:2018kmz, Hillman:2019wgh}), while facing systematically higher-order  corrections, even just at one-loop, remains an open challenge, that triggered interesting recent progress \cite{delRio:2018vrj, Chowdhury:2023khl, Beneke:2023wmt, Chowdhury:2023arc}. Higher-point functions at tree-level and their loop corrections can be quite small, and observationally undetectable within the sensibility limits of the currents experiments. Nevertheless, understanding them is of crucial importance to test the robustness of the two- and three-point tree-level predictions, along with the consistency of the theory at quantum level as well as for a general comprehension of the theory. Let us just mention that a complete understanding of the infra-red effects in the presence of sufficiently light states is still needed: large logarithms appear and accumulate over time, yielding the break-down of  perturbation theory \cite{Ford:1984hs, Antoniadis:1985pj, Tsamis:1994ca, Tsamis:1996qm, Tsamis:1997za, Polyakov:2007mm, Senatore:2009cf, Polyakov:2009nq, Giddings:2010ui, Burgess:2010dd, Marolf:2010nz, Marolf:2010zp, Krotov:2010ma, Giddings:2011zd, Marolf:2011sh, Giddings:2011ze, Senatore:2012nq, Pimentel:2012tw, Polyakov:2012uc, Senatore:2012wy, Hu:2018nxy, Gorbenko:2019rza, Mirbabayi:2019qtx, Baumgart:2019clc, Cohen:2020php, Baumgart:2020oby, Mirbabayi:2020vyt, Cohen:2021fzf, Premkumar:2021mlz, Cespedes:2023aal, Bzowski:2023nef, Benincasa:2024rfw}. From a more phenomenological standpoint, the importance of having a handle on the structure beyond the tree-level approximation arises as the coupling between inflaton and fermions appears at one-loop \cite{Green:2013rd, Chen:2016hrz, Chen:2016uwp, McAneny:2017bbv}. Furthermore, even the basic question on the types of functions to expect for a given process at lower order in perturbation theory is unanswered.

This state-of-the-art has to be contrasted with the one in {\it flat-space}
 where a number of techniques have been developed for bypassing the difficulties of the direct Feynman diagrams computations and led, in an idealized case, to the computation of a six-particle scattering amplitude at eight loops \cite{Dixon:2023kop}. Indeed, this difference is easy to understand as, in the cosmological case, the mode functions are in general given by Hankel functions which quickly complicate the computations. On a more conceptual level, the breaking of both time-translation invariance and Lorentz invariance leads to a richer analytic structure even in the simpler cases when the mode functions are related to plane-waves such as conformally-coupled scalars with polynomial interactions \cite{Arkani-Hamed:2017fdk}. 

The Bunch-Davies perturbative wavefunction associated to a graph $\mathcal{G}$ for a large class of scalar toy models in FRW cosmologies -- which include conformally-coupled scalars as well as massless and certain light states --, can be represented as an integral of the flat-space wavefunction associated to the very same graph $\mathcal{G}$ over kinematic space, whose measure encodes the details of the cosmology and the states involved \cite{Arkani-Hamed:2017fdk, Benincasa:2019vqr}. For power-law cosmologies, the map between flat-space and cosmological wavefunction is simply given by a (multiple) Mellin-transform. Such a representation allows identifying that the result of such an integration for both tree-level graphs and loop {\it integrands} can be expressed in terms of polylogarithms: this can be seen by both extracting the symbols using the geometry of the underlying {\it cosmological polytopes} \cite{Arkani-Hamed:2017fdk, Hillman:2019wgh}, and by studying, just at tree-level, the differential equations such integrals satisfy \cite{De:2023xue, Arkani-Hamed:2023bsv, Arkani-Hamed:2023kig}.

It turns out that given any regulated integral, new functions can be obtained by suitably varying the kinematic parameters, {\it i.e.} by analytic continuation, amounting to applying differential operations/variations. Owing to the twisted period integral representation \cite{Mastrolia:2018uzb} of these functions, their finite-dimensional vector space structure can be investigated by means of methods borrowed from computation algebraic and differential geometry, such as polynomial ideals algebra  \cite{Gluza:2010ws,Larsen:2015ped,Ita:2015tya,Bosma:2017hrk} and intersection theory \cite{Mastrolia:2018uzb, Frellesvig:2019kgj, Frellesvig:2019uqt, Mizera:2019gea, Frellesvig:2020qot, Caron-Huot:2021xqj,Caron-Huot:2021iev, Chestnov:2022xsy, Fontana:2023amt, Brunello:2023rpq, Brunello24}. Moreover, the dimension of this space can be computed combinatorially as the number of critical points of an appropriate  Morse's height function \cite{Lee:2013hzt, Mastrolia:2018uzb, Frellesvig:2019kgj}.
Within this context, cosmological observables can be thought as generated by a finite set of basic functions, cast into a vector $\mathcal{J}$ that satisfies a linear system of first order differential equations \cite{Gehrmann:1999as, Argeri:2007up, Henn:2013pwa, Argeri:2014qva, Lee:2014ioa}, of the form
\begin{equation}\label{eq:de1}
    \mbox{d}\mathcal{J}\:=\:
    {\mathbb A}\,\mathcal{J} \ , 
\end{equation}
where d is the total differential in kinematic space, and ${\mathbb A}$ is a matrix-valued one-form, referred to as connection matrix, generically depending on the kinematics and on the regularization parameters.

The general challenge is to find the matrix ${\mathbb A}$ and determine the possible singularities of this system of differential equations, as well as to classify the types of integrals that arise when considering the (Laurent/Taylor) series expansion of \eqref{eq:de1} around suitable, critical values of the regularization parameters - which may eventually correspond to special cosmological scenarios.

In this paper, we begin with a systematic study of the space of functions of cosmological integrals at one-loop, setting up the differential equation \eqref{eq:de1} for two prototypical examples of one-loop graphs, {\it i.e.} the two- and three-site graphs. For the first type of one-loop corrections, as for the tree-level case, the elements of $\mathcal{J}$ can be related to bounded regions in the space of integration variables determined by the hyperplanes associated to the loci where the integrand becomes singular and to those where the integration measure vanishes. These features seem not to replicate neither in the three-site case nor in any other one-loop graph, for the hyperplane arrangement of the integration measure (factorizing in a product of linear polynomials) seems to be a prerogative of just in the two-site case -- see \cite{Benincasa:2024lxe} and references therein. 

We show that the two-site graph can be expressed in terms of polylogarithms, whereas the three-site diagram -- and presumably graphs with a higher number of sites -- involves also {\it elliptic functions}. 
This complexity emerging already at the one-loop order contrasts with the simplicity of Feynman graphs in flat-space, where elliptic integrals appear at higher-loop orders -- 
the two-loop (two-dimensional) sunrise integral with massive propagators being 
the simplest graph giving rise to  integrals containing elliptic functions as well as other non-polylogarithic functions   \cite{Broadhurst:1993mw,Laporta:2004rb,Bloch:2013tra,Bloch:2014qca,Broadhurst:2016myo,Bogner:2017vim}. As a consistency check of our results, we show how the usual flat-space polylogarighmic structure emerges on the total energy conservation sheet.

The paper is organized as follows. In Section \ref{sec:cosmoint}, we review the salient features of the cosmological Feynman integrals appearing in the Bunch-Davies wavefunction. We recall the salient features of the loop measure as well as of the integrand and its definition in terms of cosmological polytopes. In particular, we briefly discuss a special class of triangulations, leading to partial fraction identities that will be used in the main body of the paper. Section \ref{sec:method} is devoted to developing the relation between cosmological integrals and differential equations. We also remark how at loop level, these integrals can be written in terms of twisted period integrals. Section \ref{sec:measure} discusses one special class of twisted integrals appearing for any one-loop graph, as well as the counting of the associated master integrals. In Section \ref{sec:Bubble}, we provide the explicit example of the one-loop two-site graph (or bubble graph), for which we study the differential equations for the loop integration. Through direct integration we instead map this result, otherwise valid only for the flat-space wavefunction and for conformal interactions in power-law FRW cosmologies, to the one-loop two-site contributions to the wavefunction for scalars with arbitrary polynomial interactions in power-law FRW cosmologies. Section \ref{sec:Triangle} discusses the cosmological integral associated to the one-loop three-site graph, where we emphasize the appearance of elliptic functions. Section \ref{sec:outlooks} contains our conclusion and outlook. Finally, the Appendices \ref{sec:A1} and \ref{sec:A2} provide further details on the relation with intersection theory and on the differential equations for the one-loop two-site graph.

%% file: sections/cosmologicalintegrals.tex
\section{\label{sec:cosmoint}Cosmological Integrals}

The probability distribution behind the correlations at late-time of an expanding universe are encoded into the so-called wavefunction of the universe (see \cite{Benincasa:2022gtd} and references therein):
\begin{equation}\label{eq:wf}
    \Psi[\Phi]\:=\:\mbox{exp}
    \left\{
        -\sum_{n\ge2}
        \int\prod_{j=1}^n
        \left[
            \frac{d^d p^{\mbox{\tiny $(j)$}}}{(2\pi)^d}\,
            \Phi(\vec{p}^{\mbox{\tiny $(j)$}})
        \right]
        \tilde{\psi}_n
        \left(
            \{\vec{p}^{\mbox{\tiny $(j)$}}\}
        \right)
    \right\}
\end{equation}
where the $n$-point wavefunction coefficient $\tilde{\psi}_n$ can be expressed, in perturbation theory, in terms of graphs
\begin{equation}\label{eq:wfg}
    \tilde{\psi}_n\:=\:\sum_{\mathcal{G}\subset\mathcal{G}_n}
    \tilde{\psi}_{\mathcal{G}} \ , 
\end{equation}
$\mathcal{G}_n$ being the collection of graphs with $n$ external states. These coefficients represent the transition amplitude from the vacuum, to a certain state configuration at late times. The vacuum is taken to be of Bunch-Davies type, {\it i.e.} the mode functions at early times select just the positive energy\footnote{As it is customary in the literature about this subject, with a little abuse of language we refer to the moduli of the momenta as {\it energies}, {\it e.g.} $E_j:=|\vec{p}^{\mbox{\tiny $(j)$}}|$.} solutions, decaying exponentially upon suitable regularization via the $i\epsilon$-prescription\footnote{For a detailed analysis of the $i\epsilon$-prescription see \cite{Albayrak:2023hie} and references therein.}

For a large class of scalar toy models, the contribution $\tilde{\psi}_{\mathcal{G}}$ to the wavefunction from a graph $\mathcal{G}$ can be represented as an integral whose integrand is independent of the cosmology -- and it is nothing but the flat-space wavefunction contribution associated to the same graph $\mathcal{G}$ --, while the details of the cosmology are encoded into the integration measure \cite{Arkani-Hamed:2017fdk, Benincasa:2019vqr}:
\begin{equation}\label{eq:wfgint}
    \tilde{\psi}_{\mathcal{G}}(X,Y)\:=\:
    \prod_{s\in\mathcal{V}}
    \left[
        \int_{X_s}^{+\infty}
        dx_s\
        \tilde{\lambda}(x_s-X_s)
    \right]
    \psi_{\mathcal{G}}(x,Y) \ , 
\end{equation}
with $\mathcal{V}$ being the set of sites of $\mathcal{G}$, $x:=\{x_s,\,s\in\mathcal{V}\}$  being the sets of weights associated to the sites, while $Y:=\{y_e,\,e\in\mathcal{E}\setminus\mathcal{E}^{\mbox{\tiny $(L)$}}\}$ are the set of edge weights associated to the edges in the tree structures within $\mathcal{G}$ -- $\mathcal{E}$ indicates the set of all the edges of $\mathcal{G}$ and $\mathcal{E}^{\mbox{\tiny $(L)$}}$ is the set of its edges in loops. The pair $(x,Y)$ parametrizes the kinematic space of the flat-space wavefunction $\psi_{\mathcal{G}}$, where $x_s$ represents the sum of the energies of the external state at the site $s$ while $y_e$ represents the total momentum flowing through the edge $e$. Finally, $X:=\{X_s,\,s\in\mathcal{V}\}$ -- together with $Y$ -- parametrizes instead the kinematics in the expanding universe, where $X_s$ has still the interpretation of the sum of energies at a site $s$: the integral \eqref{eq:wfgint} is a map from the flat-space to the FRW kinematics.

For expanding power-law cosmologies, this map is just a Mellin transform, as the integration measure, for $k$-point interaction, is simply given by \cite{Arkani-Hamed:2017fdk}
\begin{equation}\label{eq:MTm}
    \tilde{\lambda}(x)\:=\:\lambda_k\frac{i^{\alpha}}{\Gamma(\alpha)}\,x^{\alpha-1},
    \quad
    \alpha\::=\:\gamma\left[2-\frac{(d-1)(k-2)} {2}\right] \ , 
\end{equation}
where $\lambda_k$ is the coupling, and $\gamma$ encodes the cosmology\footnote{The value of the parameter $\alpha$ in equation \eqref{eq:MTm} strictly holds for conformally coupled scalars. For light scalars, such a value get shifted and depends on the states propagating in the edges as well as on the external states at a given site \cite{Benincasa:2019vqr}.} -- {\it e.g.} $\gamma=1$ returns the de Sitter space-time.

Interestingly, a similar integral representation holds for the spatial correlators -- see \cite{Benincasa:2022gtd,Benincasa:2024leu}. So, given a graph $\mathcal{G}$, contributing indistinctly to the wavefunction or to the correlators, one can talk about a cosmological integral associated to it:
\begin{eqnarray}
	&\mathcal{I}_{\mathcal{G}}[\alpha,\,\beta;\,\mathcal{X}]\sim
    \int\limits_{\mathbb{R}_+^{n_s}}
    \prod_{s\in\mathcal{V}}
    \left[
        \frac{dx_s}{x_s}\,x_s^{\alpha_s}
    \right]
    \int_{\Gamma}
    \prod_{e\in\mathcal{E}^{\mbox{\tiny $(L)$}}}
    \left[
        \frac{dy_e}{y_e}\,y_e^{\beta_e}
    \right]\times \nonumber \\
    &
    \times\mu_{d}(y_e;\mathcal{X})\,
	\frac{\mathfrak{n}_{\delta}(x,y;\,\mathcal{X})}{\displaystyle
            \prod_{\mathfrak{g}\subseteq\mathcal{G}}
	    \left[q_{\mathfrak{g}}(x,y;\,\mathcal{X})\right]^{\tau_{\mathfrak{g}}}} \, ,
\label{eq:integral}
\end{eqnarray}
where ``$\sim$'' indicates the omission of the factors depending on the couplings and on $\alpha$, $q_{\mathfrak{g}}$ are linear polynomials in $x$ and $y$ and $n_{\delta}$ is a polynomial of degree $\delta$; $\mathcal{X}$ indicates the set of rotational invariants parametrizing the kinematic space, which we denote by $(X_s,P_i)$; 
$\mu_{d}(y_e;\mathcal{X})$ is the measure of the loop integration, parameterized here by the integration over $\{y_e,\,e\in\mathcal{E}^{\mbox{\tiny $(L)$}}\}$, and is always positive inside the domain of integration $\Gamma$, vanishing only at the boundary. Note that, with respect to \eqref{eq:wfgint}, in the integral \eqref{eq:integral} the loop integration has been made explicit, and the integration over the site weights has been shifted, $x_s\longrightarrow x_s+X_s$. 

\noindent
{\bf The loop integration measure --} In general, a cosmological integral such as \eqref{eq:integral} associated to a graph $\mathcal{G}$, is a Mellin transform over the site weights and the loop edge weights, of the product of a rational function times the function $\mu$, the latter being either rational or multi-valued. Interestingly, $\mu$ is given in terms of the volume of a simplex whose edge lengths are given by the integration variables and the rotational invariants $\mu$ depends upon \cite{Benincasa:2024lxe}. The requirement that it is non-negative, together with the volumes of its faces in all codimensions, determines the contour of integration. Explicitly, the measure $\mu_d(y,\mathcal{X})$ of the edge weight integration can be written as: 
\begin{equation}\label{eq:mCM}
    \mu_d= c_{d,n_e^{\mbox{\tiny $(L)$}},L}
    \left[%
        \frac{%
            \mbox{Vol}^2
            \left\{%
                \Sigma_{n_e^{\mbox{\tiny $(L)$}}}
                \left(
                    y^2,\,P^2_{i\ldots j}
                \right)^2
            \right\}
        }{%
            \mbox{Vol}^2
            \left\{%
                \Sigma_{n_e^{\mbox{\tiny $(L)$}}-L}
                \left(
                    P^2_{i\ldots j}
                \right)
            \right\}
        }
    \right]^{\frac{d-n_s-L}{2}} \ , 
\end{equation}
and $\beta_e=2$\footnote{To be more precise, the measure \eqref{eq:mCM} is obtained proceeding loop by loop, with the contour of integration given by the intersection of the contours obtained at each loop from the non-negativity conditions mentioned in the main text. See \cite{Benincasa:2024leu} for further details.}. The powers of both numerator and denominator depend on the spatial dimension $d$, the number of sites in the loop $n_{s_l}$ and the number of loops $L$ -- see \cite{Benincasa:2024lxe} for further details. The $\Sigma$'s indicate the simplices whose dimension is reflected into its subscript, and the coefficient $c_{d,n_e^{(L)},L}$ is defined as: 
\begin{equation}\label{eq:constant}
    \begin{split}
        c_{d,n_e^{(L)},L}\:&=\:
        \frac{%
            \pi^{\frac{d-n_s+L(L-1)/2}{2}}
        }{%
            \displaystyle \prod_{l=1}^L 
            \left[
                \Gamma%
                \left(%
                    \frac{d-n_{s_l}-L+l}{2}
                \right)
            \right]
        }\,2^{n_e^{\mbox{\tiny $(L)$}}}\times\\
        &\hspace{-1cm}\times
        \left(n_e^{\mbox{\tiny $(L)$}}\right)^{d-n_e-L-2}
        \left(n_e^{\mbox{\tiny $(L)$}}!\right)
        \mbox{Vol}
        \left\{
            \Sigma_{n_e^{\mbox{\tiny $(L)$}}-L}
            \left(P^2_{i\ldots j}\right)
        \right\}.
    \end{split}
\end{equation}
\begin{figure}
    \centering
    	\includegraphics[scale=.25]{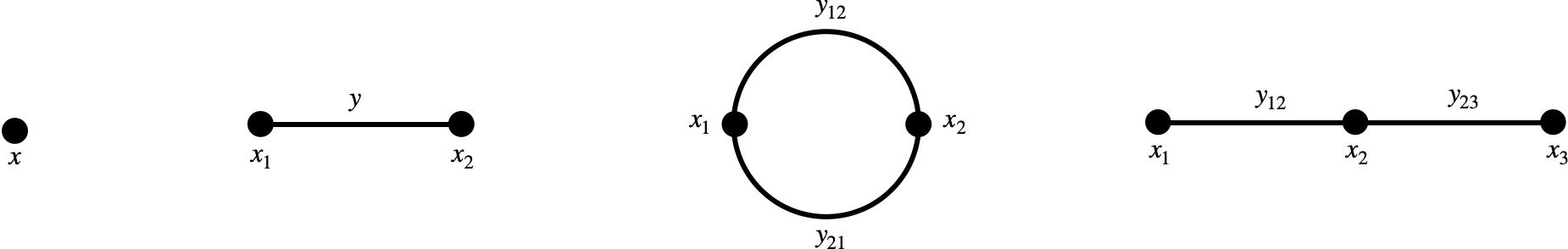}
    \caption{Graphs contributing to the wavefunction. They are characterized by weights associated to both their sites and their edges. Such weights parametrize the kinematic space for the flat-space wavefunction. The integration over the site weights and the loop edge.}
    \label{fig:graphs_exemps}
\end{figure}

\noindent
{\bf The integrand --} Given a graph $\mathcal{G}$, there is a $1-1$ correspondence between its subgraphs and the singularities of the integrand. In particular, for each subgraph $\mathfrak{g}\subseteq\mathcal{G}$, there exists a linear polynomial $q_{\mathfrak{g}}(x,y)$ defined as the sum over the weights associated to the sites in $\mathfrak{g}$ and the weights of the edges departing from it:  
\begin{equation}\label{eq:qg}
    \left\{
        q_{\mathfrak{g}}(\mathcal{Y})
        \::=\:
        \sum_{s\in\mathcal{V}_{\mathfrak{g}}}x_s\,+\,
        \sum_{e\in\mathcal{E}_{\mathfrak{g}}^{\mbox{\tiny ext}}}y_e
    ,   \,\mathfrak{g}\subseteq\mathcal{G}
    \right\},
\end{equation}
with the loci $\{q_{\mathfrak{g}}(x,y)=0,\,\mathfrak{g}\subseteq{G}\}$
identifying the poles of the integrand and determining the singularities of the integral, except for the asymptotics\footnote{For a discussion of the asymptotic behavior of the integrals \eqref{eq:integral}, see \cite{Benincasa:2024lxe}.}. For $\mathfrak{g}=\mathcal{G}$, the linear condition $q_{\mathcal{G}}:=\sum_{s\in\mathcal{V}}x_s=0$ identifies the sheet in kinematic space where the total energy of the process is conserved: as it is approached, the wavefunction reduces to the (high energy limit of the) flat-space scattering amplitudes \cite{Maldacena:2011nz, Raju:2012zr}. This limit can be used as a necessary but not sufficient condition on the consistency of the wavefunction. For any other $\mathfrak{g}\neq\mathcal{G}$, the linear condition $q_{\mathfrak{g}}=0$ identifies the kinematic sheet where the total energy of a subprocess becomes conserved: the wavefunction then factorizes into a product of (the high energy limit of) the scattering amplitude associated to the subprocess, times a linear combination of wavefunctions associated to the subgraph $\overline{\mathfrak{g}}:=\mathcal{G}\setminus\mathfrak{g}$ which differ among each other for the signs of the energies associated to the edge connecting $\mathfrak{g}$ and $\overline{\mathfrak{g}}$.

The numerator $\mathfrak{n}_{\delta}$ is a polynomial of degree $\delta$ that instead reflects the higher codimension singularity structure \cite{Benincasa:2020aoj, Benincasa:2021qcb}. As all these loci are given by homogeneous equations in the site- and edge-weights, both the site and edge weights can be taken as a set of local coordinates $\mathcal{Y}:=(x,y)$ in projective space $\mathbb{P}^{n_s+n_e-1}$. The linear polynomials can then be written as $q_{\mathfrak{g}}(\mathcal{Y})=\mathcal{Y}^{I}\mathcal{W}^{\mbox{\tiny $(\mathfrak{g})$}}_I$ where $\mathcal{W}^{\mbox{\tiny $(\mathfrak{g})$}}$ is a vector in the dual space of $\mathbb{P}^{n_s+n_e-1}$, which is still indicated as $\mathbb{P}^{n_s+n_e-1}$ -- we will refer to it as a co-vector. Considering that all the weights are positive valued, then the  linear polynomials $\{q_{\mathfrak{g}}(\mathcal{Y}),\,\mathfrak{g}\subseteq\mathcal{G}\}$ are positive. The  inequalities $\{q_{\mathfrak{g}}(\mathcal{Y})\ge0,\,\mathfrak{g}\subseteq\mathcal{G}\}$ turn out to carve out a convex space, named {\it cosmological polytope} and indicated as $\mathcal{P}_{\mathcal{G}}$ \cite{Arkani-Hamed:2017fdk}. Each co-vector $\mathcal{W}^{\mbox{\tiny $(\mathfrak{g})$}}$ identifies a hyperplane that intersect the cosmological polytope along its boundary only, containing a facet ({\it i.e.} a codimension-$1$ face) $\mathcal{P}_{\mathcal{G}}\cap\mathcal{W}^{\mbox{\tiny $(\mathfrak{g})$}}$. Given a cosmological polytope $\mathcal{P}_{\mathcal{G}}\subset\mathbb{P}^{n_s+n_e-1}$ there is a unique rational function $\Omega\left(\mathcal{Y},\mathcal{P}_{\mathcal{G}}\right)$ -- up to an overall constant --, named {\it canonical function} associated to it, which can be identified as the wavefunction universal integrand and can be written as
\begin{equation}\label{eq:CanFun}
    \Omega
    \left(
        \mathcal{Y},\mathcal{P}_{\mathcal{G}}
    \right)
    \:=\:
    \frac{%
        \mathfrak{n}_{\delta}(\mathcal{Y})
    }{%
        \displaystyle
        \prod_{\mathfrak{g}\subseteq\mathcal{G}}q_{\mathfrak{g}}(\mathcal{Y})
    }\, .
\end{equation}
which it can be readily identified in equation \eqref{eq:integral}. The denominators are associated to the facets, while the numerator is the locus of the intersection of the hyperplanes containing the facets {\it outside} $\mathcal{P}_{\mathcal{G}}$: it provides the compatibility conditions for the sequential residues \cite{Benincasa:2020aoj, Benincasa:2021qcb}. The degree of the numerator is fixed by projectivity: the differential form defined by equipping the canonical function \eqref{eq:CanFun} with the canonical measure of $\mathbb{P}^{n_s+n_e-1}$, {\it i.e.} $\langle\mathcal{Y}d^{n_s+n_e-1}\mathcal{Y}\rangle$, is invariant under $\mbox{GL}(1)$ rescaling, fixing the degree $\delta$ of $\mathfrak{n}_{\delta}$ in terms of the number of facets $\tilde{\nu}$ of $\mathcal{P}_{\mathcal{G}}$ and the dimension of projective space: $\delta=\tilde{\nu}-n_s-n_e$. The locus $\mathfrak{n}_{\delta}=0$ is the {\it adjoint} of $\mathcal{P}_{\mathcal{G}}$, and its {\it maximal subspaces} identified by the intersections of the hyperplanes containing the facets of $\mathcal{P}_{\mathcal{G}}$ outside $\mathcal{P}_{\mathcal{G}}$ such that {\it no vertex} of $\mathcal{P}_{\mathcal{G}}$ lies on the intersection, can be used to triangulate $\mathcal{P}_{\mathcal{G}}$ via a collection of polytopes $\{\mathcal{P}_{\mathcal{G}}^{\mbox{\tiny $(i)$}},\,i=1,\ldots,m\}$ whose elements' facets are all facets of $\mathcal{P}_{\mathcal{G}}$ \cite{Benincasa:2021qcb}. In terms of the canonical function, this reflects into its partial fractioning into the sum of the canonical functions of all the elements of the collection $\{\mathcal{P}_{\mathcal{G}}^{\mbox{\tiny $(i)$}},\,i=1,\ldots,m\}$, without introducing spurious poles\footnote{Recall that the singularities are associated to the facets of the polytope: if the facets of each $\mathcal{P}_{\mathcal{G}}^{\mbox{\tiny $(i)$}}$ are facets of $\mathcal{P}_{\mathcal{G}}$, then each of the terms of the partial fraction shows only singularities which are also singularities of the wavefunction.}:
\begin{equation}\label{eq:SignTr}
    \begin{split}
        \Omega
        \left(
            \mathcal{Y},\,\mathcal{P}_{\mathcal{G}}
        \right)
        \:&=\:
        \sum_{j=1}^m
        \Omega
        \left(
            \mathcal{Y},\,\mathcal{P}_{\mathcal{G}}^{\mbox{\tiny $(j)$}}
        \right)
        \:=\\
        &=\:
        \prod_{\mathfrak{g}\in\mathfrak{G}_{\circ}}
        \left[
            \frac{1}{q_{\mathfrak{g}}\left(\mathcal{Y}\right)}
        \right]
        \sum_{%
            \{\mathfrak{G}_c\}
        }
        \prod_{\mathfrak{g}'\in\mathfrak{G}_c}
        \left[
            \frac{1}{q_{\mathfrak{g}'}(\mathcal{Y})}
        \right] \ , 
    \end{split}
\end{equation}
where $\mathfrak{G}_{\circ}$ is one of the maximal subspaces of the adjoint, $\mathfrak{G}_c$ is a set of compatible graphs -- {\it i.e.} graphs that identifies singularities such that the sequential residue of the canonical function along them is non-zero -- disjoint from $\mathfrak{G}_{\circ}$, and the sum runs over all possible of these subsets. Such triangulations, triangulate a cosmological polytope using an external lower-dimensional hyperplane, identified by the intersections of the codimension-$1$ hyperplanes corresponding to the elements of $\mathfrak{G}_{\circ}$, without introducing spurious boundaries. Triangulations that only use the vertices of $\mathcal{P}_{\mathcal{G}}$, introduce spurious boundaries, translating into spurious poles in the partial fraction of the canonical function \cite{Arkani-Hamed:2017fdk, Benincasa:2018ssx, Juhnke-Kubitzke:2023nrj}.

%% file: sections/method.tex
\section{\label{sec:method}Integral Relations and Differential equations}

Let us focus on the integrals \eqref{eq:integral}, and more specifically on the integration over the edge weights:
\begin{equation}\label{eq:IntsG}
    \begin{split}
        &\mathcal{I}_{\mathcal{G}}
        \left[
            \alpha,\beta,\mathcal{X}
        \right]
        \:=\:
        \int\limits_{\mathbb{R}_+^{n_s}}
        \prod_{s\in\mathcal{V}}
        \left[
            \frac{dx_s}{x_s}\,x_s^{\alpha_s}\,
            \frac{i^{\alpha_s}}{\Gamma(\alpha_s)}
        \right]
        \mathcal{I}_{\{\tau_{\mathfrak{g}}\}}^{\mbox{\tiny $(\mathcal{G})$}}
        \left[
            \beta;\,x;\,\mathcal{X}
        \right],\\
        &\mathcal{I}_{\{\tau_{\mathfrak{g}}\}}^{\mbox{\tiny $(\mathcal{G})$}}
        :=
        \int_{\Gamma}
        \prod_{e\in\mathcal{E}^{\mbox{\tiny $(L)$}}}
        \left[
            \frac{dy_e}{y_e}\,y_e^{\beta_e}
        \right]
        \mu_d\left(y_e;\mathcal{X}\right)\,
        \frac{\mathfrak{n}_{\delta}(x,y;\mathcal{X})}{%
            \displaystyle
            \prod_{\mathfrak{g}\subseteq\mathcal{G}}
            \left[
                q_{\mathfrak{g}}(x,y;\mathcal{X})
            \right]^{\tau_{\mathfrak{g}}}
        } 
    \end{split}
\end{equation}
Interestingly, the integrals $\mathcal{I}_{\{\tau_{\mathfrak{g}}\}}^{\mbox{\tiny $(\mathcal{G})$}}$ can be recast in terms of {\it twisted period integrals}
\begin{equation}\label{eq:y_integral}
    \mathcal{I}_{\{\tau_{\mathfrak{g}}\}}^{\mbox{\tiny $(j)$}}:= \int_\Gamma u\, \varphi \   \ , 
    \qquad 
    \varphi:= 
    \frac{%
        \displaystyle
        \prod_{e\in\mathcal{E}^{\mbox{\tiny $(L)$}}}dy_e}{%
        \displaystyle 
        \prod_{\mathfrak{g}\in
        \mathfrak{G}^{\mbox{\tiny $(j)$}}\cup\{e\}} 
        \left[
            q_{\mathfrak{g}}(y)
        \right]^{\tau_{\mathfrak{g}}}
    },
\end{equation} 
where $\mathfrak{G}^{\mbox{\tiny $(j)$}}$ represents a collection of $j$ subgraphs of $\mathcal{G}$, while $\{e\}$ indicates a subset of the edges of $\mathcal{G}$ such that $q_e(y)=y_e$. Furthermore, 
\begin{equation}\label{eq:ukchi}
    \begin{split}
        &u\: :=\:\mu_d\:=\: \kappa_0 \, \bk^\gm, \qquad
            \bk\: :=\mbox{Vol}^2
                \left\{%
                    \Sigma_{n_e^{\mbox{\tiny $(L)$}}}
                    \left(%
                        y^2,P^2_{i\ldots j}
                    \right)
                \right\} \ , \\ 
    &\gm\::=\:\frac{d-n_s-L}{2},\\
    &\kappa_0\::=\:c_{d,n_e^{(L)},L}
        \left[
            \mbox{Vol}^2
            \left\{
                \Sigma_{n_e^{\mbox{\tiny $(L)$}}-L}
                \left(%
                    P_{i\ldots j}^2
                \right)
            \right\}
        \right]^{-\gm}, 
    \end{split}
\end{equation}
with $\bk$ playing the role of a {\it twist}, since it vanishes on the integration boundary:
\begin{equation}
   \bk|_{\partial \Gamma} = 0 \ . 
   \label{eq:twist}
\end{equation}
This property ensures that integrals of the same family can be seen as bilinear pairings between differential $n$-forms $\varphi$, considered as an element of the $n$-th de-Rham cohomology group for twisted cocycles (defined as the vector space of closed forms modulo exact forms) \cite{Mastrolia:2018uzb,Frellesvig:2019uqt,Frellesvig:2020qot}, and the integration domain $\Gamma$, considered as an element of the  $n$-th de-Rham homology group for twisted cycles (defined as the vector space of closed $n$-chains  modulo boundary terms). 
Such vector spaces are isomorphic and finite-dimensional, and their dimension $\nu$ corresponds to the number of independent forms (or independent contours), and equivalently, to the number of independent integrals of the family, referred to as master integrals. Introducing the regularized twist $u_{\rho}$ and 
the connection $\omega_\rho$, defined as :
\begin{equation}\label{eq:u}
    \begin{split}
        &u_{\tau}= u(y) 
        \prod_{\mathfrak{g}\in\mathfrak{G}^{\mbox{\tiny $(j)$}}\cup\{e\}} q_{\mathfrak{g}}^{\tau_{\mathfrak{g}}} \ , 
        \\
        &\omega_{\tau}= d\log(u_\tau) = \sum_{e\in\mathcal{E}_{\mathfrak{g}}}\omega_e^{\mbox{\tiny $(\tau)$}} dy_e \, ,  
    \end{split}
\end{equation}  
where each $\tau_{\mathfrak{g}}$ plays the role of an analytic regulator, the dimension $\nu$ can be computed as the number of critical points of the function $\log(u_{\tau})$ \cite{Lee:2013hzt,Mastrolia:2018uzb,Frellesvig:2019uqt}, determined as:
\begin{eqnarray}
    \nu\  :=\  \text{number of solutions of}\  \omega_{\tau}=0 \ \ , 
    \label{eq:counting}
\end{eqnarray}
where $\omega_{\tau}=0$ is a zero dimensional system of linear equations in the variables ${\bf y}$.

Within the integral vector space, every integral of the type  \eqref{eq:y_integral} can be expressed as a linear combination of $\nu$ master integrals $\{\mathcal{J}_i\}_{i=1}^\nu$ as:
\begin{eqnarray}
    \mathcal{I} =\sum_{i=1}^\nu c_i \ \mathcal{J}_i \ , \qquad \mathcal{J}_i = \int_\Gamma u\  e_i.
    \label{eq:ibpdec}
\end{eqnarray}
with $c_i$ rational coefficients depending on the kinematics $\{z\in\mathcal{X}\}$ and on $\gm$, hence, on the space-time dimension $d$.

The master integrals obey linear systems of first order differential equations in the kinematic variables, 
\begin{eqnarray}
    d \mathbfcal{J} =
    d{\mathbb A} \, \mathbfcal{J} \ , \quad {\rm with} \quad
    d{\mathbb A} &=& 
    \sum_i {\mathbb A}_{z_i} \, d z_i \, \ ,
    \label{eq:DEQ1}
\end{eqnarray}
with $z_i\in \mathcal{X}$, 
where the matrix ${\mathbb A}_{z_i}$ generically depends on $z_i$'s and on $d$.

Linear relations among twisted period integrals can be obtained either using integration-by-parts identities \cite{Chetyrkin:1981qh, Laporta:2000dsw}, or via intersection numbers \cite{Mastrolia:2018uzb} -- see Appendix \ref{sec:A1}. In particular, the linear polynomials $\{q_{\mathfrak{g}},\,\mathfrak{g}\subseteq\mathcal{G}\}$ in the denominator in $\mathcal{I}_{\tau_{\mathfrak{g}}}^{\mbox{\tiny $(\mathcal{G})$}}$ are in general non-linear independent of each other: their number $\tilde{\nu}$ is higher (or equal) to the number of the variables they depend upon. More geometrically, they identify a number $\tilde{\nu}$ of co-vectors $\mathcal{W}^{\mbox{\tiny $(\mathfrak{g})$}}\in\mathbb{P}^{n_s+n_e-1}$ larger or equal to the dimensionality of the space they live in: $\tilde{\nu}\ge n_s+n_e$. This implies that $n_s+n_e$ of them, that are linearly independent, can be taken as a basis for $\mathbb{P}^{n_s+n_e-1}$ and the others can be expressed as a linear combination of them. As the co-vectors $\mathcal{W}^{\mbox{\tiny $(\mathfrak{g})$}}$ identify the facets of a cosmological polytope $\mathcal{P}_{\mathcal{G}}$, such linear combinations are fixed in its geometry, and it is precisely what the signed triangulations \eqref{eq:SignTr} make manifest and implement. When considering the full integral $\mathcal{I}_{\mathcal{G}}$ in \eqref{eq:integral}, then the signed triangulations \eqref{eq:SignTr} provides novel representations for $\mathcal{I}_{\mathcal{G}}$, that in the Feynman integral literature are referred to as {\it partial fraction identities}: They reduce the problem to a set of simpler integrals that can be studied separately.

When just the integral over the loop edge weights is considered, by considering the integrals of the form $\mathcal{I}_{\{\tau\}}^{\mbox{\tiny $(j)$}}$ in \eqref{eq:y_integral}, then further linear relations among the linear polynomials $\{q_{\mathfrak{g}},\,\mathfrak{g}\subseteq\mathcal{G}\}$ can be found: such polynomials can be still be thought as co-vectors but now living in a smaller space $\mathbb{P}^{n_e^{\mbox{\tiny $(L)$}}}$. In this case, the integrand is no-longer a canonical function of a polytope, but rather a {\it covariant function} associated to a polytope subdivision of an underlying polytope \cite{Benincasa:2020uph}. An in-depth discussion of this point of view is beyond the scope of this paper. What is important to highlight here is that, as happens for the usual cosmological polytope picture, the geometry of such polytopes and their polytope subdivisions underlies the linear relations among the $\{q_{\mathfrak{g}},\,\mathfrak{g}\subseteq\mathcal{G}\}$ seen just as functions of the loop edge weights. Schematically:
\begin{equation}\label{eq:qglincomb}
    q_{\mathfrak{g}'}(y)\,=\,c_{\mathfrak{g}'0}+
    \sum_{\tilde{\mathfrak{g}}\in\mathfrak{G}_{\mbox{\tiny B}}}
    c_{\mathfrak{g}'\tilde{\mathfrak{g}}}\,q_{\tilde{\mathfrak{g}}}(y).
\end{equation}
where $\mathfrak{G}_{\mbox{\tiny B}}$ is the set of graphs contributing to the chosen basis, $c_{\mathfrak{g}'\mathfrak{g}}$ are rational functions of the external kinematic invariants. Such linear relations generate partial fraction identities of the form
\begin{equation}
    \mathcal{I}^{\mbox{\tiny $(j)$}}_{\tau_{\mathfrak{g'}}-1}\:=\:
    c_{\mathfrak{g'}0}
    \mathcal{I}_{\{\tau_{\mathfrak{g}}\}}^{\mbox{\tiny $(j)$}}+
    \sum_{\tilde{\mathfrak{g}}\in\mathfrak{G}_{\mbox{\tiny B}}}
    c_{\mathfrak{g}'\tilde{\mathfrak{g}}}
    \mathcal{I}_{\tau_{\tilde{\mathfrak{g}}}-1}^{\mbox{\tiny $(j)$}} \ , 
\end{equation}
with $\mathcal{I}_{\{\tau_{\mathfrak{g}}\}}^{\mbox{\tiny $(j)$}}$ as in \eqref{eq:y_integral}, while $\mathcal{I}_{\tilde{\tau_{\mathfrak{g}}}-1}^{\mbox{\tiny $(j)$}}$ indicates that the integral depends on all the very same parameters $\{\tau_{\mathfrak{g}},\,\mathfrak{g}\in\mathfrak{G}^{\mbox{\tiny $(j)$}}\cup\{e\}\}$ with just the specific parameter $\tau_{\tilde{\mathfrak{g}}}$ shifted by $-1$.
\\

\noindent
{\bf Linear Algebra from Algebraic Geometry --} The vanishing of the twist on the integration boundary (equation \eqref{eq:twist}) ensures integrals of the same family to satisfy integration-by-parts identities: 
\begin{eqnarray}
  \sum_{e'\in\mathcal{E}^{\mbox{\tiny $(L)$}}} 
  \int_\Gamma 
  \left[
    \prod_{e\in\mathcal{E}^{\mbox{\tiny $(L)$}}}
    dy_e
  \right]
  \frac{\partial }{\partial y_{e'}}
  \left[ 
    \bk^\gm \,
	\frac{n_e(x,y,\mathcal{X})}{%
        \displaystyle
        \prod_{g\subseteq\mathcal{G}}
        \left[
            q_{\mathfrak{g}}
        \right]^{\tau_{\mathfrak{g}}}
	}
   \right]
   = 0  \ , 
     \label{eq:IBP}
\end{eqnarray}
where $n_{e}(x,y;\,\mathcal{X})$ are generic numerators depending on the integration variables and on the external kinematics.

Upon the action of the derivative in \eqref{eq:IBP}, 
the emerging relations may involve integrals with different powers of $\bk$, namely living in shifted space-time dimensions, 
unless the (free) numerators $n_a$ are chosen to fulfill the condition
~\cite{Gluza:2010ws,Larsen:2015ped,Ita:2015tya,Bosma:2017hrk,DGPS}:
\begin{eqnarray}
    \sum_{e\in\mathcal{E}^{\mbox{\tiny $(L)$}}} 
    \left(
        \partial_{y_e}\bk
    \right) \, n_e = n_0 \, \bk \ .
\label{eq:condition}
\end{eqnarray}
The latter can be viewed as an equation for the polynomials $n_e$ and $n_0$, which can be determined a solution of \textit{Syzygy}  equations  within the module  $\langle \bk,\partial_{y_1}\bk,\ldots, \partial_{y_n}\bk \rangle \, $. 

By row-reducing the system \eqref{eq:IBP} \cite{Peraro:2016wsq,Peraro:2019svx}, loaded with \textit{Syzygy} equations, it is possible to find an independent set of $\nu$ independent master integrals $\{\mathcal{J}_i\}_{i=1}^\nu$, 
that constitute an integral basis, 
for decomposing every integral of the type \eqref{eq:y_integral}, as in equation \eqref{eq:ibpdec}.

Homogeneous systems of differential equations satisfied by the master integrals of type  \eqref{eq:DEQ1} can be obtained by decomposing the derivative of the master integrals with respect to a kinematic variable $x$ in terms of master integrals:
\begin{equation}
    \partial_x \, \mathcal{J}_i = \sum_j ({\mathbb A}_{x})_{ij} \, \mathcal{J}_j \ . 
    \label{eq:generic_deq}
\end{equation}
In this work, by following a procedure similar to the one adopted for the generation of the integration by parts identities, also for the generation of the differential equations, relations involving integrals with
different powers of $\bk$ can be eliminated by requiring that $\partial_x \, \bk$ is an element of the ideal $\langle \bk,\partial_{y_1}\bk,\ldots, \partial_{y_n}\bk \rangle \, $, namely by requiring that there exist polynomials in $y_e$ (but rational functions in the kinematic variables), $\{w_i\}_{i=0}^{n_e^{\mbox{\tiny $(1)$}}}$, such that:
\begin{eqnarray}
    \partial_x  \bk = 
    \sum_{i=1}^{n_e^{\mbox{\tiny $(1)$}}} 
    w_i \, \partial_{y_i} \bk 
    + w_0 \, \bk \, .
    \label{eq:DEQ_syzygy}
\end{eqnarray}
The functions $w_i$ and $w_0$  can be conveniently built from ratio of polynomials, say  $r_i$ $(i=0,\ldots, n+1)$, found  as solution of \textit{Syzygy} equations,
 \begin{eqnarray}
    r_{n+1} \, \partial_x  \bk = 
    \sum_{i=1}^n 
    r_i \, \partial_{y_i} \bk 
    + r_0 \, \bk \, ,
\end{eqnarray}
having $w_i \equiv r_i/r_{n+1}$, where $r_{n+1}$ is chosen to be independent of $\{y_e,\,e\in\mathcal{E}^{\mbox{\tiny $(1)$}}\}$.

\noindent
{\bf Linear Algebra from Intersection Theory -- } Within the intersection theory approach proposed in  \cite{Mastrolia:2018uzb},
the master integrals can be chosen a priori, soon after determining the dimension $\nu$ of the vector space.
In this case, alternatively to the row reduction algorithm mentioned in the earlier paragraph,
the integral decomposition in terms of master integrals, as well as the generation of the system of differential equations the latter obey, can be directly obtained simply as a {\it projection} operation in a vector space, 
computing the coefficients $c_i$ of equation \eqref{eq:ibpdec}, and the elements of the matrix  ${\mathbb{A}_x}$ of equation \eqref{eq:generic_deq},
in terms of {\it intersection numbers} \cite{Frellesvig:2019uqt,Frellesvig:2019kgj,Frellesvig:2020qot,Chestnov:2022xsy,Brunello:2023rpq,Brunello24}. Details are provided in Appendix \ref{sec:A1}.

%% file: sections/Fullintegration.tex
\noindent
{\bf Integrals from Differential Equations -- }
In general, finding an exact solution of the differential equations \eqref{eq:DEQ1} is a challenging task. However, the system can be solved around fixed spacetime dimension, setting $d=3+2\epsilon$, with $\epsilon$ playing the role of an analytic regulator, which is possible in the case of conformally coupled scalars. \\
One can choose a set of master integrals $\mathbfcal{J}$  that obeys a {\it canonical} system of differential equations \cite{Henn:2013pwa,Argeri:2014qva,Lee:2014ioa}, 
\begin{eqnarray}
\label{eq:canonicaldeq}
    d \mathbfcal{J} = \epsilon \ d \hat{{\mathbb A}}\ \mathbfcal{J}\ ,  \qquad d =  \sum_{x\in \mathcal{X}} dx\ \partial_x \ , 
\end{eqnarray}
where  the $\epsilon$-dependence is factorised from the kinematic variables, and the total differential matrix $d \hat{{\mathbb A}}$ is a logarithmic differential form.
The solution can be formally written as:
\begin{eqnarray}
    \mathbfcal{J} = \mathbb{P} \, \exp\left(\epsilon\int_\gamma d \hat{\mathbb{A}}\right)\mathbfcal{J}_0, 
    \label{eq:path_order}
\end{eqnarray}
where the path order exponential operator, playing the role of an evolution operator, acts on the vector $\mathbfcal{J}_0 \equiv 
\mathbfcal{J}\big|_{\partial \gamma}$, {\it i.e.} the vector of master integrals evaluated at the boundary of the integration domain. This vector encodes the initial conditions,
hereafter simply referred to as boundary conditions. 
The latter have to be provided independently, and, in general, their values can be obtained from the knowledge of simpler integrals, or by imposing regularity conditions.
The advantage of the canonical form is that the solution of the differential equation can be expanded as a Taylor series in $\epsilon$, and expressed in terms of iterated integrals \cite{Chen:1977oja}:
\begin{eqnarray}
    \mathbfcal{J} = \left(\mathbb{I} + \epsilon \int_\gamma d \hat{\mathbb{A}} +\frac{\epsilon^2}{2!}\int_\gamma d \hat{\mathbb{A}} \int_\gamma  d \hat{\mathbb{A}}+\ldots\right) \mathbfcal{J}_0
    \  .
\end{eqnarray}

{\it Picard-Fuchs --} In order to understand the class of iterated integrals that appear, it is useful to study systems of differential equations sector-by-sector, where an integral sector $(\sigma_1,\ldots,\sigma_{\tilde{\nu}})$ is defined as the set of points $(\tau_{\mathfrak{g}})_{\mathfrak{g}\subseteq\mathcal{G}} \in\mathbb{Z}^{\tilde{\nu}}$ such that $\sigma_i = \theta(\tau_i-1/2)$. Differential equations have a block-triangular structure, whose homogeneous sector univocally determines the space of functions appearing in its solution. In particular, each homogeneous block of dimension $k \times k$ of the system in equation \eqref{eq:DEQ1} can be equivalently rewritten as a single $k^{\rm th}$-order differential equation for a single master integral $\mathcal{J}_i$, as:
\begin{equation}
    \mathcal{L}_k \mathcal{J}_i = 0 \ , \qquad \mathcal{L}_k = \frac{\partial^k}{\partial x^k} + \sum_{j=0}^{k-1} c_j(x)\ \frac{\partial^j}{\partial x^j} \ , 
    \label{eq:picard-fuchs}
\end{equation}
with $c_j(x)$ rational functions in the kinematic variable $x$.
In the multivariate case, given $\{x_i\}_{i=1}^m$ kinematic variables, this can be achieved by first making the change of variables $x_i\to a_i \lambda$, setting $a_k=1$, and then constructing the higher order operator with respect to $\lambda$.
This operator is known as Picard-Fuchs operator \cite{Muller-Stach:2012tgj}, and by studying its factorization property it is possible to determine the geometries of the appearing integrals \cite{Muller-Stach:2012tgj,Adams:2017tga}: if it factorizes in differential operators of order 1, only polylogarithms \cite{Remiddi:1999ew, Goncharov:1998kja} appear, for $r>1$ instead we can have the appearance of Calabi-Yau ($r$-1)-folds \cite{Broedel:2017siw,Broadhurst:1993mw,Broedel:2018iwv}.

{\bf Site weight integration --} According to what outlined, the evaluation of the cosmological integrals of type ${\cal I}_{\cal G}$, defined in \eqref{eq:integral}, may be carried out by adopting a two-fold strategy, 
where: in the first phase, the innermost integrals over the $y_e$ variables are expressed as a linear combination of canonical master integrals; 
and, in the second phase, the integration over the $x_s$ variables is performed by means of Mellin transforms.

%% file: sections/PartialFractioning.tex
\section{Integrals over the one-loop measure}\label{sec:measure}

In what follows, we will be mostly concerned with one-loop integrals. In particular, we will be interested in those integrals purely constituted by a loop, {\it i.e.} with no tree substructure. They correspond to polygons, with the number of their sites and edges which coincide. As it is therefore possible to identify such graphs by the pair $(n_s,1)$ constituted by the number of sites and loops ($L=1$) respectively, we find it convenient to introduce the replacement $\mathcal{G}\longrightarrow(n_s,1)$ in the notation for the integrals, so that the integrals in the two lines in \eqref{eq:IntsG} are now given by $\mathcal{I}_{\mathcal{G}}\longrightarrow\mathcal{I}_{\mbox{\tiny $(n_s,1)$}}$ and $\mathcal{I}^{\mbox{\tiny $(n_s,1)$}}_{\{\tau_{\mathfrak{g}}\}}$ respectively\footnote{Note that without the restriction to purely loop graphs, the pair $(n_s,1)$ would not uniquely identify a graph. For example, for $n_s=3$, there exists also a one-loop three-site graph constituted by a one-loop two-site subgraph having one of its sites connected to a third site by an edge. This restriction happens without loss of generality, as the ``excluded'' graphs contain smaller polygons and hence are included in the discussion.}.

Before diving into specific one-loop integrals, we will discuss the integral family associated just to the integration measure, 
in the case where no denominators is present, namely considering integrals of the type:
\begin{equation}
    \mathcal{I}^{\mbox{\tiny $(n_s,1;0)$}}_{\{1\}} \: :=\:\int_{\Gamma}
    \prod_{e\in\mathcal{E}^{\mbox{\tiny $(1)$}}} 
    \left[dy_e\right] \; \mu_d(y;\,\mathcal{X})\, \ = \ \kappa_0 \int_{\Gamma} \bk^\gamma(y)\ , 
    \label{eq:zero_cosmo}
\end{equation}
where $\mu_d$ is given in equation \eqref{eq:mCM}. 
One can relate the integration measure of one-loop $n$-points cosmological integrals, with that obtained from the Baikov polynomial \cite{Baikov:1996iu,Frellesvig:2017aai} of one-loop $n$-points Feynman integrals with different masses on external legs and with massless propagators, in Euclidean space,
by considering $\{y_e^2,\,e\in\mathcal{E}^{\mbox{\tiny $(1)$}}\}$. This property is relevant, because the number of master integrals for these Feynman integral families admits a recursive formula, given by: 
\begin{eqnarray}
    \nu_{n}^{\mbox{\tiny (FI)}} \ = \ \sum_{k=2}^n \frac{n!\ }{k!\ (n-k)!}  \ = \ 2^n - (n+1)\ ,
\end{eqnarray} 
where exactly 1 master integral per sector appears.
Integration by parts identities will shift powers of denominators by integer units, relating integrals on different sub-sectors.
Rewriting equation \eqref{eq:zero_cosmo} in Baikov variables, 
one obtains a family of one-loop $n$-points Feynman integrals with variables raised to half-integer powers:
\begin{eqnarray}
    \mathcal{I}^{\mbox{\tiny $(n_s,1;0)$}}_{\{1\}} = 
    \frac{\kappa_0}{2^n} \int \ 
    \frac{%
        \displaystyle
        \prod_{e\in\mathcal{E}^{\mbox{\tiny $(1)$}}} dy_e^2}{%
        \displaystyle
        \prod_{e\in\mathcal{E}^{\mbox{\tiny $(1)$}}} 
        \left[y_e^2\right]^{1/2}
    } \,  \left[\bk(y_e^2)\right]^\epsilon \ ,
    \label{eq_zero_sec}
\end{eqnarray}
which in momentum space correspond to:
\begin{align}    
   \mathcal{I}^{\mbox{\tiny $(n_s,1;0)$}}_{\{1\}}
        =
    \frac{1}{2^n}\int_{\mathbb{R}^n}d\vec{l}\,
        \frac{1}{%
            \displaystyle
            \left|\vec{l}\right| 
            \left|\vec{l}+\vec{P}_1\right|\cdots  
            \left|\vec{l}+\sum_{j=1}^{n_s-1}\vec{P}_j\right|
        } \, .
    \label{eq:mom_n_0}
\end{align}
In general, such integral belongs to the integral family: 
\begin{align}
    \mathcal{I}_{\{\tau_{\mathfrak{g}}\}}^{\mbox{\tiny $(n_s,1;0)$}}
      &   :=
        \int \ 
    \frac{%
        \displaystyle
        \prod_{e\in\mathcal{E}^{\mbox{\tiny $(1)$}}}
        dy_e^2
    }{%
        \displaystyle
        \prod_{e\in\mathcal{E}^{\mbox{\tiny $(1)$}}} \left(y_e^2\right)^{\tau_e} 
    } \,  \left[\bk(y^2)\right]^\epsilon \ \\
    & = 
    \int_{\mathbb{R}^{n_s}}d\vec{l}\, 
        \frac{1}{[(\vec{l})^2]^{\tau_{12}} \cdots  [(\vec{l}+\vec{P}_1+\ldots +\vec{P}_{n_s-1})^2]^{\tau_{n_s,1}}} 
    \label{eq:subsec_n_0}
\end{align}
with $\tau_e\in \mathbb{Z}+1/2$, $e\in\mathcal{E}^{\mbox{\tiny $(1)$}}$. Integrals of the type of equation \eqref{eq_zero_sec} cannot be related to subsectors where some denominators do not appear, and for each master Feynman integral appearing with $k$ external legs, we have a sector with $\nu_{k}^{\mbox{\tiny (FI)}}$ master cosmological integrals. The total number of master integral of the zero sector is:
\begin{eqnarray}
    \nu_{n_s}^{\mbox{\tiny (CI)}} \ = \ 
    \sum_{k=2}^{n_s} \frac{n_s!\ }{k!(n_s-k)!}\  
    \nu_k^{\mbox{\tiny (FI)}} \ ,
\end{eqnarray}
where the various subsectors are appearing in the with differential equations blocks of dimension $\nu_k^{\mbox{\tiny (FI)}}$.
Summing the series, we obtain:
\begin{eqnarray}
    \nu_{n_s}^{\mbox{\tiny (CI)}} \ = \  3^{n_s}-2^{n_s-1}(2+n_s)  \ .
\end{eqnarray} 

%% file: sections/applications.tex
\section{\label{sec:Bubble}One-loop two-site graph}
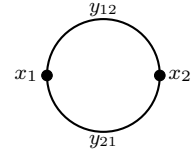
\begin{figure}[t]
	\centering
	\begin{tikzpicture}[ball/.style = {circle, draw, align=center, anchor=north, inner sep=0}, 
		cross/.style={cross out, draw, minimum size=2*(#1-\pgflinewidth), inner sep=0pt, outer sep=0pt}]
	  	\begin{scope}
			\coordinate[label=left:{\footnotesize $x_1$}] (x1) at (0,0);
			\coordinate[label=right:{\footnotesize $x_2$}] (x2) at ($(x1)+(1.5,0)$);
			\coordinate (c) at ($(x1)!.5!(x2)$);
			\draw[thick] (c) circle (.75cm);
			\draw[fill] (x1) circle (2pt);
			\draw[fill] (x2) circle (2pt);
			\node[scale=.875] at ($(c)+(0,.875)$) {$y_{12}$};
			\node[scale=.875] at ($(c)-(0,.875)$) {$y_{21}$}; 
		\end{scope}	
	\end{tikzpicture}
	\caption{One-loop two-site diagram. The corresponding integrand has 5 denominators, each corresponding to a connected subgraph of the above graph. Two subgraphs enclose each of the two sites, then two subgraphs enclose both sites and cut one edge twice and finally there is the full graph which corresponds to the total energy pole.}
	\label{fig:Bubble}
\end{figure}

In this section, we discuss how all the technology outlined in the previous sections manifests itself in the simplest case of the one-loop two-site integral and allows us to get insights on the integrated function, -- see Figure \ref{fig:Bubble}. 

Let us consider the following representation for this integral:
\begin{eqnarray}
        \mathcal{I}_{\mbox{\tiny $(2,1)$}}&=&
           \int_{\mathbb{R}^2_+}\prod_{s\in\mathcal{V}} 
           \left[
                \frac{dx_s}{x_s}\,x_s^{\alpha}
            \right]
        \mathcal{I}_{\{1\}}^{\mbox{\tiny $(2,1)$}} \ , 
         \label{eq:x_bubble}\\
        \mathcal{I}_{\{1\}}^{\mbox{\tiny $(2,1)$}} &= &
            \kappa_0\int_{\Gamma
            }\prod_{e\in\mathcal{E}^{\mbox{\tiny $(1)$}}}
            \left[dy_{e}\,y_e\right]
           \,\frac{\bk^{\gm}}{\q_{\mathcal{G}} \q_{\mathfrak{g}_1} \q_{\mathfrak{g}_2}}\left(\frac{1}{\q_{\mathcal{G}_{12}}}+\frac{1}{\q_{\mathcal{G}_{21}}} \right) \ \ \ 
    \label{eq:full_bubble}
\end{eqnarray}
where $\mathcal{E}^{\mbox{\tiny $(1)$}}:=\{e_{12},e_{21}\}$ is the set of the two edges connecting the sites $s_1$ and $s_2$, and $\bk$, $\kappa_{\circ}$ and $\gm$ can be obtained from \eqref{eq:ukchi} by setting $L=1$, $n_e^{\mbox{\tiny $(1)$}}=n_s=2$,
while $\mathfrak{g}_j$ identifies the subgraph containing just the site $s_j$ (whose weight is $x_j+X_j$) and $\mathcal{G}_{ij}:=\mathcal{G}\setminus\{e_{ij}\}$ is the subgraph obtained from $\mathcal{G}$ by deleting the edge between the sites $s_i$ and $s_j$ -- in this simple case where there are only two sites, the two edges are indicated by reversing the order of the labels of the sites they connect. The linear polynomials associated to these subgraphs are given by \ref{eq:qg}, which can be explicitly written as,
\begin{equation}
    \begin{split}
        &\q_{\mathcal{G}} = \tilde{x}_1+\tilde{x}_2, \\
        &\q_{\mathfrak{g}_1} = \tilde{x}_1+y_{12} + y_{21} ,\\
        &\q_{\mathfrak{g}_2} = \tilde{x}_2+y_{12} + y_{21} ,\\
        &\q_{\mathcal{G}_{12}} = \tilde{x}_1+\tilde{x}_2+2y_{12},\\ 
        &\q_{\mathcal{G}_{21}} = \tilde{x}_1+\tilde{x}_2+2y_{21}.
    \end{split}
\end{equation}
where for simplicity we denoted $\tilde{x}_i = x_i+X_i$.

{\bf Loop edge weight integration --} Upon exploiting the invariance of the integrand under the $y_{112}\leftrightarrow y_{21}$ exchange, and the partial fraction relations emerging from the identity:
\begin{equation}
    \q_{\mathfrak{g}_1} - \q_{\mathfrak{g}_2} = \tilde{x}_1-\tilde{x}_2 \ , 
\end{equation}
the integral $\mathcal{I}_{\{1\}}^{\mbox{\tiny $(2,1)$}}$,
appearing in \eqref{eq:full_bubble},
can be recast as a combination of twisted period integrals \eqref{eq:y_integral}, corresponding to two sets of two denominators, namely $\{ \q_{\mathfrak{g}_1}, \q_{\mathcal{G}_{12}} \}$ and  $\{\q_{\mathfrak{g}_2} , \q_{\mathcal{G}_{21}}\}$. The latter set can be mapped onto the former, by exchanging $\tilde{x}_1 \leftrightarrow \tilde{x}_2$. Therefore, the computational complexity of the problem reduces remarkably to the evaluation of just one type of period integrals, defined as:
\begin{equation}
    {\cal I}_{\tau_{\mathfrak{g}_1} \tau_{\mathcal{G}_{12}}}^{\mbox{\tiny $(2,1)$}} := \int_\Gamma  \bk^\gm \, 
    \varphi_{\tau_{\mathfrak{g}_1} \tau_{\mathcal{G}_{12}}}\ , \ \ \  \varphi_{\tau_{\mathfrak{g}_1} \tau_{\mathcal{G}_{12}}} := \frac{dy_{12}dy_{21}}{\q_{\mathfrak{g}_1}^{\tau_{\mathfrak{g}_1}}\q_{\mathcal{G}_{12}}^{\tau_{\mathcal{G}_{12}}}}  \ .
    \label{eq:2site}
\end{equation}

From equation \eqref{eq:counting} it is possible to identify a number $\nu = 6$ of master integrals, which can be chosen as: $\mathbfcal{I}=\{\mathcal{I}_{00},\mathcal{I}_{10},\mathcal{I}_{01},\mathcal{I}_{02},\mathcal{I}_{-11},\mathcal{I}_{11}\}$ -- since now on we suppress the superscript for notational convenience.

As described in Section \ref{sec:measure}, the sector without denominators contains $\nu_2^{\mbox{\tiny{(CI)}}}=1$ master integrals, which has been chosen as $\mathcal{I}_{00}$, and in momentum space it can be rewritten as a massless one-loop two point function with massive external momenta, belonging to the integral family $\mathcal{I}_{\tau_{\mathfrak{g_1}}\tau_{\mathfrak{g}_2}}$, with denominators raised to with half-integer exponents, in Euclidean space:
    \begin{equation}
        \mathcal{I}_{\tau_{\mathfrak{g_1}}\tau_{\mathfrak{g}_2}} := \int \frac{d\vec{\ell}}{[(\vec{\ell})^2]^{\tau_{\mathfrak{g}_1}}[(\vec{\ell} +\vec{P})^2]^{\tau_{\mathfrak{g}_2}}}\, , \ \ \tau_{\mathfrak{g}_j}\in \mathbb{Z}+\frac{1}{2} 
       \ .  \label{eq:flat_bubble}
     \end{equation}
Using the algebraic geometry methods of section \ref{sec:method}, and independently using intersection theory \cite{Mastrolia:2018uzb}, as shown in appendix \ref{sec:A1}, it is possible to obtain the differential equations obeyed by the master integrals.
With a change of basis $\mathbfcal{J} = \mathcal{R}.\mathbfcal{I}$, through the rotation matrix $\mathcal{R}$ described in appendix \ref{sec:A2}, it is possible to find a family of master integrals:
 
\begin{align}
& &\mathcal{J}_1 =\,  &\frac{(1+2 \epsilon)^2}{P^2} \, \mathcal{I}_{00} \ ,& \nonumber \\ 
& &\mathcal{J}_2 =\,  &\frac{\epsilon (1+2 \epsilon)  }{P} \, \mathcal{I}_{10} \ , &\nonumber  \\
& &\mathcal{J}_3 =\, & \frac{1}{P}\bigl(\epsilon (1+4 \epsilon) \, \mathcal{I}_{01}
+ \epsilon(\tilde{x}_1+\tilde{x}_2) \, \mathcal{I}_{02}\bigr) \ , &\nonumber\\ 
& & \mathcal{J}_4 =\,  &- \epsilon  \ \mathcal{I}_{02} \ , &\nonumber \\ 
& & \mathcal{J}_5 =\,  &\frac{\epsilon (1+2\epsilon)}{2P (\tilde{x}_1+\tilde{x}_2)} \bigl(  \mathcal{I}_{-11} - \mathcal{I}_{00} + (\tilde{x}_2-\tilde{x}_1) \, \mathcal{I}_{01}
    \bigr)\ , &   \nonumber \\
&& \mathcal{J}_6  =\, & \epsilon^2 \, \mathcal{I}_{11}    \ ,& 
\end{align}
obeying a canonical system of differential equations \cite{Henn:2013pwa,Argeri:2014qva,Lee:2014ioa}, as  defined in \eqref{eq:canonicaldeq}, 
where the total differential matrix,
\begin{eqnarray}
 d \mathbb{A} &=& 
   \hat{\mathbb{A}}_{\tilde{x}_1} d\tilde{x}_1
 + \hat{\mathbb{A}}_{\tilde{x}_2} d\tilde{x}_2 
 + \hat{\mathbb{A}}_{P} \, dP \nonumber\\
 &=& \sum_{i=1}^{8} \mathbb{M}_i \, d\log (w_i) \ , 
 \label{eq:can_deq_1L}
\end{eqnarray}
is in $\dd \log$ form: $\mathbb{M}_i$ are constant matrices whereas the letters $w_i \in \{P,\tilde{x}_1+\tilde{x}_2, \tilde{x}_1 + P, \tilde{x}_2 + P , \tilde{x}_1 + \tilde{x}_ 2 + 2P, \tilde{x}_1 - P , \tilde{x}_2 - P, \tilde{x}_1 + \tilde{x}_ 2 - 2P \}$ (the last three entries correspond to spurious singularities) form a rational alphabet. The system of differential equations admits a solution in terms of iterated integrals, as  shown in \eqref{eq:path_order}, which in this case give rise to generalized polylogarithms \cite{Goncharov:1998kja,Remiddi:1999ew,Vollinga:2004sn,Duhr:2019tlz}. The analytic expression for our master integrals up to order $\mathcal{O}(\epsilon^2)$ is obtained after fixing boundary conditions either via direct integration or imposing regularity at the spurious singularity - see appendix \ref{sec:A2}, for details. 
Using the results of the master integrals, $\mathcal{I}_{\{1\}}^{\mbox{\tiny $(2,1)$}}$ reads as,
\begin{eqnarray}
\mathcal{I}_{\{1\}}^{\mbox{\tiny $(2,1)$}} &=  &
-\frac{1}{\epsilon (\tilde{x}_1+\tilde{x}_2)}+\frac{  (-2 \log (P)-\gamma_E +2-\log (4 \pi ))}{\tilde{x}_1+\tilde{x}_2} \nonumber\\
& & \nonumber \\ 
& +& \frac{2}{\tilde{x}_1^2-\tilde{x}_2^2} \left[\tilde{x}_2 \log \left(\frac{P+\tilde{x}_1}{P}\right)-\tilde{x}_1 \log \left(\frac{P+\tilde{x}_2}{P}\right)\right] \nonumber\\
& & \nonumber \\ 
   & -&\frac{1}{P}  \biggl[\frac{\pi ^2}{6}+\text{Li}_2\left(\frac{P-\tilde{x}_2}{P+\tilde{x}_1}\right)+ \text{Li}_2\left(\frac{P-\tilde{x}_1}{P+\tilde{x}_2}\right)\nonumber\\
& & \nonumber \\ 
   & +&\frac{1}{2} \log
   ^2\left(\frac{P+\tilde{x}_1}{P+\tilde{x}_2}\right)\biggr] \ . 
\end{eqnarray}    
{\bf Site weight integration --} The integration over the $x$-variables of equation \eqref{eq:x_bubble} can be performed directly in terms of known Mellin transforms \cite{mellinbook}, and via the Method of Brackets \cite{Gonzalez:2010zzb,Ananthanarayan:2021not}. Such method is based on Ramanujan's master theorem which states that given a complex valued function $g(x)$, which can be Taylor expanded around $x\to 0$ as:

\begin{equation}
g(x)=\sum_{k=0}^{\infty} \frac{G(k)}{k !}(-x)^k \ , 
\end{equation}
then its Mellin transform is given by
\begin{equation}
\int_0^{\infty} x^{s-1} g(x) d x=\Gamma(s) G(-s) \ . 
\end{equation}
The final result, which is symmetric under the exchange of $X_1\leftrightarrow X_2$, can be expressed as a linear combination of \textit{Hypergeometric} functions ${}_2F_1$ and ${}_3F_2$ and logarithms, and reads as:

\begin{widetext}
\begin{small}
\begin{eqnarray}
\mathcal{I}_{\mbox{\tiny $(2,1)$}} &=& 
  \frac{2^{-3 - 2\alpha} \pi^{3/2} (X_1 + X_2)^{1 + 2\alpha} \csc(\pi \alpha)^2 
   \Gamma\left(-\frac{1}{2} - \alpha\right) 
   }{\Gamma[-\alpha]}\left(2 - \frac{1}{\epsilon} 
   - \log(4\pi e^{\gamma_E} P^2 ) 
  \right) \nonumber \\
& & 
+\frac{\pi ^{3/2} \csc ^2(\pi  \alpha ) }{8 (\alpha \
+1)^2 P}\
\biggl[
-4 \sqrt{\pi } \left((P+X_1)^{\alpha +1}-2 \
(X_1-P)^{\alpha +1}\right) (P+X_2)^{\alpha +1}
\nonumber \\
& & 
-\frac{4^{-\alpha } \Gamma \left(-\alpha \
-\frac{1}{2}\right) (X_1+X_2)^{2 \alpha +2} \, }{\Gamma \
(-\alpha )}{}_2F_1\left(1,-2 (\
\alpha +1);-\alpha ;\frac{P+X_1}{X_1+X_2}\right)
\biggr] \nonumber \\
&&
+\frac {\pi ^2 \csc (\pi  \alpha ) \csc (2 \pi  \alpha ) (P + X_ 1)^{\alpha }} {4 \alpha + 2} \biggl[-2 (P + X_ 1) \left((P - X_ 2)^{\alpha } + (-1)^{\alpha } (P + X_ 2)^{\alpha } \right)
\nonumber 
\end{eqnarray}
\end{small}
\end{widetext}
\newpage
\begin{widetext}
\begin{small}
\begin{eqnarray}
\qquad \qquad \qquad  \qquad 
& & 
+(-1)^{\alpha } (X_ 1 - X_ 2) (P + 
     X_ 1)^{\alpha } \, _ 2 F_ 1\left(1 - \alpha , -2 \alpha ; 
     1 - 2 \alpha ;\frac {X_ 1 - X_ 2} {P + X_ 1} \right) \nonumber \\
& & 
+ (X_ 1 + 
    X_ 2) (P + 
      X_ 1)^{\alpha } \, _ 2 F_ 1\left(1 - \alpha , -2 \alpha ; 
    1 - 2 \alpha ;\frac {X_ 1 + X_ 2} {P + X_ 1} \right)
      \biggr] \nonumber \\
   &&
   -\frac {\pi ^{5/2} 4^{-\alpha - 
       1} \csc (\pi  \alpha ) \csc (2 \pi  \alpha )} {\Gamma (-\alpha \
) \Gamma \left (\alpha + \frac {3} {2} \right) (P + X_ 1)}\biggl[
(-1)^{\alpha } (X_1-X_2)^{2 \alpha +2} \, _3F_2\left(1,1,\alpha +2;2,2 \
\alpha +3;\frac{X_1-X_2}{P+X_1}\right)\nonumber \\
& & 
+(X_1+X_2)^{2 \alpha +2} \, \
_3F_2\left(1,1,\alpha +2;2,2 \alpha +3;\frac{X_1+X_2}{P+X_1}\right)
\biggr]
\nonumber \\
&&
+\frac{\pi ^{5/2} 2^{-2 \alpha -1} \csc (\pi  \alpha ) \csc (2 \pi  \
\alpha )  \left((-1)^{\alpha } \
(X_1-X_2)^{2 \alpha +1}+(X_1+X_2)^{2 \alpha +1}\right)}{\Gamma \
(-\alpha ) \Gamma \left(\alpha +\frac{3}{2}\right)}\log \left(\frac{P+X_1}{P}\right) \nonumber \\
&& + (X_1 \leftrightarrow X_2).
\end{eqnarray}
\end{small}
\end{widetext}
\par In the limit $\alpha \to 1$, the above answer has at most transcendental weight two functions.

\section{\label{sec:Triangle}One-loop three-site graph}

\begin{figure}[t]
	\centering
	\begin{tikzpicture}[ball/.style = {circle, draw, align=center, anchor=north, inner sep=0}, 
		cross/.style={cross out, draw, minimum size=2*(#1-\pgflinewidth), inner sep=0pt, outer sep=0pt}]
		\begin{scope}[shift={(4.75,0)}, transform shape]
			\coordinate[label=above:{\footnotesize $x_1$}] (x1) at (0,.75);
			\coordinate[label=below:{\footnotesize $x_2$}] (x2) at (-.875,-.75);
			\coordinate[label=below:{\footnotesize $x_3$}] (x3) at (+.875,-.75);
			\draw[thick] (x1) --node[midway, left, scale=.875] {$y_{12}$} (x2) --node[midway, below, scale=.875] {$y_{23}$} (x3) --node[midway, right, scale=.875] {$y_{31}$} cycle;
			\draw[fill] (x1) circle (2pt);
			\draw[fill] (x2) circle (2pt);
			\draw[fill] (x3) circle (2pt);
		\end{scope}
	\end{tikzpicture}
	\caption{One loop three-site diagram. The corresponding integrand has 10 denominators, each corresponding to a connected subgraph of the above graph. Three subgraphs which enclose a single site, three which enclose two sites at a time, three which enclose all three sites but cut each edge twice and finally the full graph which corresponds to the total energy singularity.}
	\label{fig:Triangle}
\end{figure}
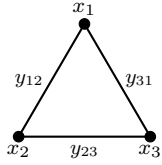

Let us move on to the next-to-simplest one-loop case, constituted by the one-loop three-site integral -- See Fig. \ref{fig:Triangle}.  As we will show, it has some distinctive features which were absent in the previous case. The easiest to spot is the fact that now the volumes in the edge weight measure are higher degree polynomials that no longer factorizes in a product of linear polynomials. Actually, such a factorization occurs for the one-loop two-site case only.

In what follows, we restrict ourselves to the case in which there is just one external state at each site, so that 
$|\vec{P}_i| \to X_i$. Reducing the number of scales from six to three simplifies the problem while still capturing all the essential complexities. 

The representation for the integrand coming from one of the sign triangulations \ref{eq:SignTr} of the underlying cosmological polytope, which corresponds to the choice $\mathfrak{G}_{\circ}=\{\mathcal{G},\,\mathfrak{g}_1,\,\mathfrak{g}_2,\,\mathfrak{g}_3\}$, is given in terms of the sum of six simplices. Interestingly, it is enough to focus on the study of the differential equations for one of them, as the others can be derived through permutations of integration variables and external kinematics. Explicitly, such representation for the integrand yields the following form for the integral $\mathcal{I}_{\{1\}}^{\mbox{\tiny $(3,1)$}}$:
\begin{equation}\label{eq:Triangle}
    \begin{split}
        \mathcal{I}_{\{1\}}^{\mbox{\tiny $(3,1)$}}
        &=
        \kappa_0\int_{\Gamma}\,
        \prod_{e\in \mathcal{E}^{\mbox{\tiny $(1)$}}} 
        \left[
            dy_e\,y_{e}
        \right]\,
        \frac{\bk^{\gm}}{%
            \displaystyle
            \q_{\mathcal{G}} \prod_{j=1}^3\q_{\mathfrak{g}_j} 
        }\times
        \\
        &\times
        \left[%
            \frac{1}{\q_{\mathcal{G}_{12}}}
            \left(%
                \frac{1}{\q_{\mathfrak{g}_{23}}}+
                \frac{1}{\q_{\mathfrak{g}_{31}}} 
            \right)+
        \right.\\
        & 
        \left.
            +\frac{1}{\q_{\mathcal{G}_{23}}}
            \left(%
                \frac{1}{\q_{\mathfrak{g}_{31}}}+
                \frac{1}{\q_{\mathfrak{g}_{12}}}
            \right)+
            \frac{1}{\q_{\mathcal{G}_{31}}}
            \left(%
                \frac{1}{\q_{\mathfrak{g}_{12}}}+
                \frac{1}{\q_{\mathfrak{g}_{23}}}
            \right) 
        \right]
    \end{split}
\end{equation}
where $\bk$, $\gm$ and $\kappa_{\circ}$ can be obtained from \ref{eq:ukchi} by setting $L=1$ and $n_s=3$.
Furthermore, for regularization purposes, we can consider $d=3+2\epsilon$. The set of edges $\mathcal{E}^{\mbox{\tiny $(1)$}}$ is given by $\mathcal{E}^{\mbox{\tiny $(1)$}}:=\{e_{12},\,e_{23},\,e_{31}\}$. Finally, it is useful to write here the explicit expression for the linear polynomials $\{q_{\mathfrak{g}},\,\mathfrak{g}\subseteq\mathcal{G}\}$, whose associated subgraphs follow the same conventions introduced in the previous section with $\mathfrak{g}_{s_1\ldots s_{\tilde{n}_s}}$ being the connected subgraph containing the sites $s_1,\ldots\,s_{\tilde{n}_s}$, while $\mathcal{G}_{ij}:=\mathcal{G}\setminus\{e_{ij}\}$ is the subgraph obtained from $\mathcal{G}$ by deleting the edge $e_{ij}$ connecting the sites $s_i$ and $s_j$: 
\begin{equation}
    \begin{split}
        & \q_{\mathcal{G}} = \sum_{i=1}^3 X_i,\\
        & \q_{\mathfrak{g}_j} = y_{j-1,j}+X_j+y_{j,j+1},\\
        & \q_{\mathcal{G}_{j,j+1}} = \sum_{s=1}^3 X_s+y_{j,j+1} \ , 
    \end{split}
\end{equation} 
with $j=1,2,3$. 

Partial fraction identities allows focusing only on subsets of three denominators: the evaluation can then be split into two types of contributions,  separating the calculation of the sectors with denominators $\{\q_{\mathfrak{g}_j},\,j=1,2,3\}\cup\{\q_{\mathfrak{g}_{23}}\}$, and of ones containing $q_{\mathcal{G}_{12}}$ and the pairs $\{(q_{\mathfrak{g}_j},\,q_{\mathfrak{g}_{j+1}}),\,
(q_{\mathfrak{g}_j},\,q_{\mathfrak{g}_{24}});\,j=1,2,3\}$. 
\\

\begin{figure}[t]
	\centering
	 \includegraphics[scale=0.2]{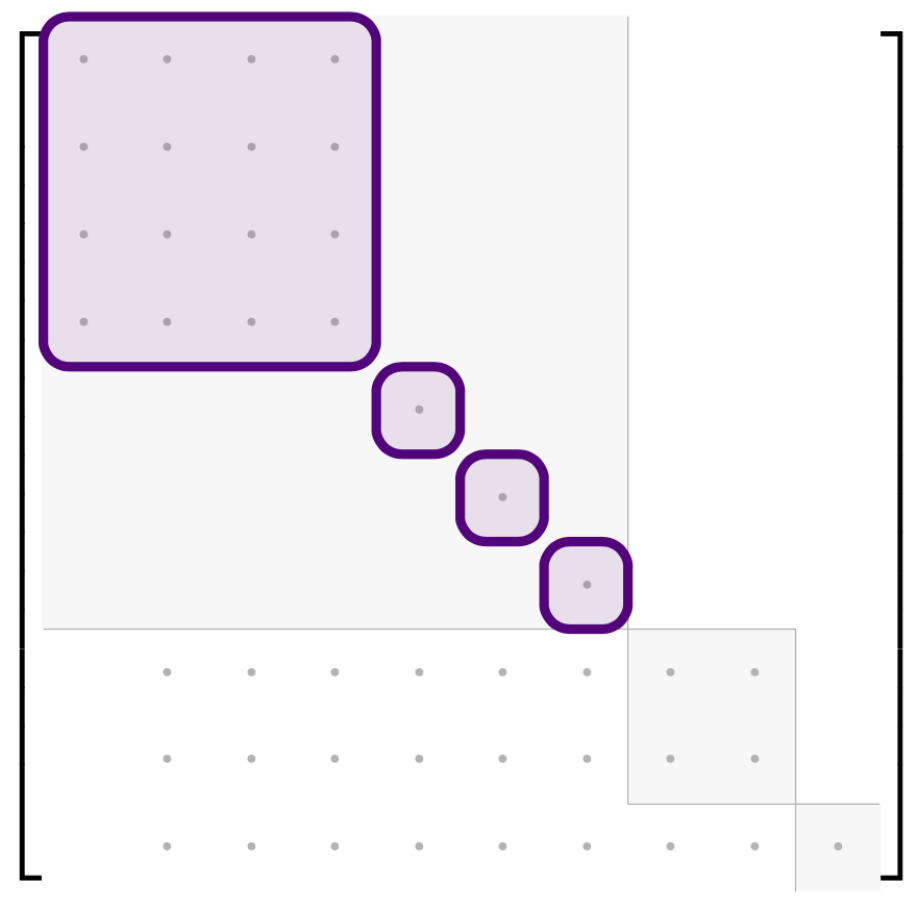}
	\caption{Zero sector of the one-loop three-site-graph. The first four integrals form a $4\times 4$ homogeneous diagonal block, corresponding to the one-loop three point function with denominator raised to half-integer powers, and each of the subsequent three integrals form a $1\times 1$ diagonal block, and can be identified with one-loop two-point functions with denominators raised to half integer powers.}
	\label{fig:zerosec}
\end{figure}

{\bf Polylogarithmic sector --} Let us begin with the sector identified by $\{\q_{\mathfrak{g}_j},\,j=1,2,3\}\cup\{\q_{\mathfrak{g}_{23}}\}$. The associated integrals can be written as
\begin{equation}\label{eq:subsec_1235}
    \begin{split}
    &\mathcal{I}_{%
        \tau_{\mathfrak{g}_1}\tau_{\mathfrak{g}_2}\tau_{\mathfrak{g}_3}
        \tau_{\mathfrak{g}_{23}}
    } 
    \:=\: 
    \int_\Gamma \,\mu_d\, 
    \varphi_{\tau_{\mathfrak{g}_1}\tau_{\mathfrak{g}_2}\tau_{\mathfrak{g}_3}\tau_{\mathfrak{g}_{23}}} \ , 
    \\ 
    &\varphi_{%
        \tau_{\mathfrak{g}_1}\tau_{\mathfrak{g}_2}\tau_{\mathfrak{g}_3}
        \tau_{\mathfrak{g}_{23}}
    }
    \:=\:
    \frac{%
        \displaystyle
        \prod_{e\in\mathcal{E}^{\mbox{\tiny $(1)$}}}
        dy_e
    }{%
        \q_{\mathfrak{g}_{1}}^{\tau{\mathfrak{g}}_1} 
        \q_{\mathfrak{g}_{2}}^{\tau{\mathfrak{g}}_2}
        \q_{\mathfrak{g}_{3}}^{\tau{\mathfrak{g}}_3}
        \q_{\mathfrak{g}_{23}}^{\tau_{\mathfrak{g}_{23}}} 
    } \ . 
    \end{split}
\end{equation}
From the counting procedure prescribed by equation \eqref{eq:counting},  the integral family has $15$ master integrals, whose master forms can be chosen as:
\begin{align}
    &e_1=\varphi_{0000}\ ,&\, 
    &e_2=y_{12}^2 \varphi_{0000}\ ,& \, 
    &e_3=y_{23}^2 \varphi_{0000} \ ,& \, \nonumber \\ 
   &  e_4=y_{31}^2 \varphi_{0000}\ ,&\,
    &e_5=y_{12} \varphi_{0000}\  ,&\, 
   & e_6=y_{23} \varphi_{0000}\  ,&\, \nonumber\\
  &  e_7=y_{31} \varphi_{0000}\  ,& \, 
     & e_8=\varphi_{1000}\  ,&\, 
   & e_9=\varphi_{0100}\ ,&\, \nonumber \\
   & e_{10}=\varphi_{0010}\  ,&\, 
  &  e_{11}=y_{23}\varphi_{0001}\  ,&\, 
    &e_{12}=y_{31}\varphi_{0001} \ ,&\,\nonumber \\ 
&    e_{13}=y_{12}\varphi_{0001} \ ,&\,
 &   e_{14}=\varphi_{0002} \ ,&\,
 &   e_{15}=\varphi_{1110} \ .\,&
 \label{eq:triangle_mis}
\end{align}
As described in section \ref{sec:measure}, the sub-sector without denominators contains $\nu_3^{\mbox{\tiny{(CI)}}}=7$ master integrals,
and its differential equations are shown in figure \ref{fig:zerosec}. Heuristically, this large number can be motivated by rewriting the measure of the integral in momentum space, which belongs to the integral family:
\begin{align}
        \mathcal{I}^{\mbox{\tiny $(3,1;0)$}}_{%
            \tau_{1}\tau_{2}\tau_{3}
        }
        =
        \int_{\mathbb{R}^3}d\vec{l}\,
        \frac{1}{[(\vec{l})^2]^{\tau_1} 
        [(\vec{l}+\vec{P}_1)^2]^{\tau_2} 
        [(\vec{l}+\vec{P}_1+\vec{P}_2)^2]^{\tau_3}} \, ,
    \label{eq:subsec_1235_0}
\end{align}
where $\tau_i\in \mathbb{Z}+1/2$.
The integral in equation \eqref{eq:subsec_1235_0} is the one-loop three-point function with massive external momenta of mass $P_i$ and with massless denominator raised to half-integer powers, in Euclidean spacetime.
Integration by parts in $y_e$ will mix integrals with denominators raised to half-integer powers with those raised to integer ones, a property that does not hold for momentum space integration by parts identities, and which effectively increases the number of master integrals.

It is possible to find a $\epsilon$-factorized form for the differential equation matrices obeyed by these integrals \cite{Argeri:2014qva,Meyer:2017joq,Dlapa:2020cwj,Frellesvig:2021hkr,Dlapa:2022wdu}.
Also in this case, the total differential can be written in \textit{dlog} form, indicating that the space of functions consists of generalized polylogarithms, and the alphabet for this sector, together with the equivalent sectors in the remaining three similar integrals (obtained by replacing $\q_{\mathfrak{g}_{23}}$ with $\q_{j,j+1}$, $j\neq2$), reads:
\begin{align}
  W \ = \   \{X_1, X_2, X_3,X_1+X_2, X_2 + X_3, X_1 + X_3,\nonumber \\
    -X_3 + X_1 - X_2, X_3 + X_1 - X_2,  \nonumber \\
    -X_3 + X_1 + X_2, X_3 + X_1 + X_2\}\, .
\end{align}

\par 
In the generic case of multiple external legs, in which $x_i\neq P_i$, the basis of this sector increases to $34$. The function space consists only of generalized polylogarithms, but algebraic letters appear in the alphabet.
\\

{\bf Elliptic sector --} Let us now turn our attention to the sectors containing the denominator $\q_{\mathcal{G}_{12}}$:
\begin{equation}\label{eq:elliptic_subsec}
    \begin{split}
        &\mathcal{I}_{%
            \tau_{\mathfrak{g}}\tau_{\mathfrak{g}'}
            \tau_{\mathcal{G}_{12}}
        } 
        \:=\:
        \int_\Gamma \,\mu_d \, 
        \varphi_{%
            \tau_{\mathfrak{g}}\tau_{\mathfrak{g}'}
            \tau_{\mathcal{G}_{12}}},\\
        &\varphi_{\tau_{\mathfrak{g}}\tau_{\mathfrak{g}'}
            \tau_{\mathcal{G}_{21}}}
        \:=\:
        \frac{%
            \displaystyle
            \prod_{e\in\mathcal{E}^{\mbox{\tiny $(1)$}}}dy_e
        }{%
        \q_{\mathfrak{g}}^{\tau_{\mathfrak{g}}}
        \q_{\mathfrak{g}'}^{\tau_{\mathfrak{g}'}}
        \q_{\mathcal{G}_{12}}^{\tau_{\mathcal{G}_{12}}} } ,
    \end{split}
\end{equation}
where $(\mathfrak{g},\mathfrak{g}')$ takes values in the set of pairs $\{(\mathfrak{g}_j,\mathfrak{g}_{j+1}),\,(\mathfrak{g}_j,\,\mathfrak{g}_{23})\}$.

\begin{figure}[t]
	\centering
	 \includegraphics[scale=0.12]{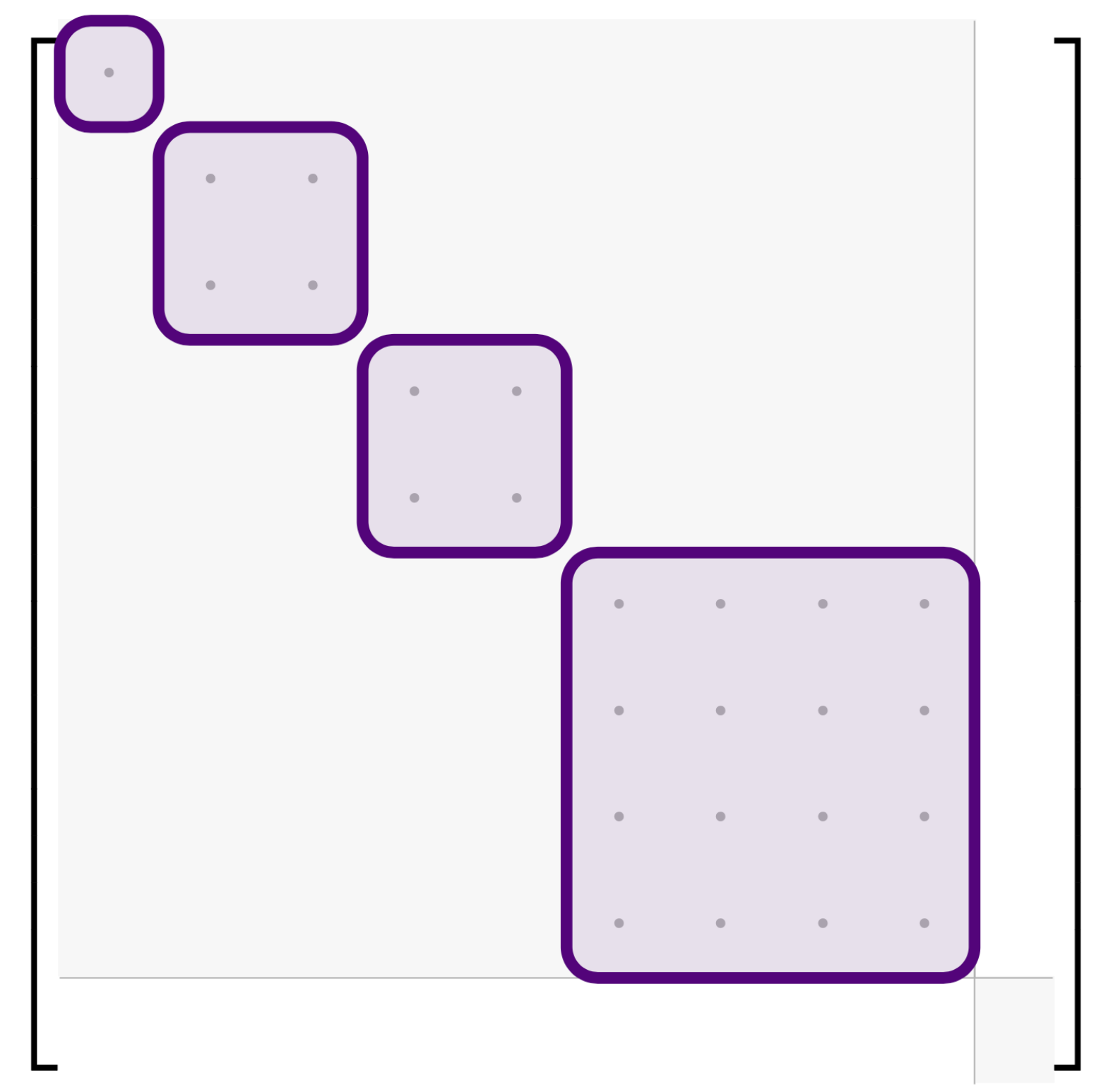}
	\caption{Homogeneous sector of the one-loop three-site graph with denominator $\q_a$. There are a total of 9 master integrals, which decouple in blocks of dimensions 
 $1 \times 1,
  2 \times 2,
  2 \times 2,
  4 \times 4$. 
 In the last block, the elliptic family appears.}
	\label{fig:elliptic_sec}
\end{figure}

The sub-sector containing only the denominator $\q_{\mathcal{G}_{12}}$, has $9$ master integrals, that can be chosen as follows:
\begin{align}
    &e_1= y_{23} y_{31}  \varphi_{001}\ ,&\, 
    &e_2=y_{23}  \varphi_{001}\ ,& \, 
    &e_3=y_{23} \varphi_{002} \ ,& \, \nonumber \\ 
   &  e_4=y_{31} \varphi_{001}\ ,&\,
    &e_5=y_{31} \varphi_{002}\  ,&\, 
   & e_6= \varphi_{002}\  ,&\, \nonumber\\
  &  e_7= \varphi_{001}\  ,& \, 
     & e_8= y_{23}^2 \varphi_{001}\  ,&\, 
   & e_9= y_{31}^2 \varphi_{001}\ ,&\, 
\label{eq:elliptic_sector}
\end{align}
and whose shape of the differential equation is shown in Fig.\ref{fig:elliptic_sec}.
 Constructing the Picard-Fuchs operators for each homogeneous block of the differential equation in $d=3$ as in equation \eqref{eq:picard-fuchs}, where we used the change of variables: $X_1 \to a_1 \lambda, \ X_2 \to \lambda,\  X_3\to 1$, we found a differential operator of third order $\mathcal{L}_3$, corresponding to the sector formed by the last 3 master integrals of equation \eqref{eq:elliptic_sector}, which factorizes \cite{maple} in one operator of first order and one of second order, as $\mathcal{L}_3=\mathcal{L}_1 \mathcal{L}_2$.
The operator $\mathcal{L}_2$ reads:
\begin{align}
   & \mathcal{L}_2= \frac{d^2}{d\lambda^2}+ \frac{5 \left(a^2-1\right)^2 \lambda ^4-6 \left(a^2+1\right) \lambda ^2+1}{\left(a^2-1\right)^2 \lambda ^5-2 \left(a^2+1\right) \lambda
   ^3+\lambda }\frac{d}{d\lambda} \nonumber\\
   & +\frac{3 \left(a^2-1\right)^2 \lambda ^2-2 \left(a^2+1\right)}{\left(a^2-1\right)^2 \lambda ^4-2 \left(a^2+1\right) \lambda ^2+1}  \ , 
   \label{eq:PF_operator}
\end{align}
and its solution is an elliptic function:
\begin{equation}
\psi_{1,2}(K^2)\ , \ \ \ K^2 = \frac{(a-1)^2 \lambda ^2-1}{(a+1)^2 \lambda ^2-1} \ . 
\end{equation}
Remarkably, the space of function gets richer already at one-loop level with respect to the Feynman integral case.
As a consistency check of our result, we expect the elliptic function to disappear in the flat spacetime limit, in which $\q_{\mathcal{G}}\to 0$. With the change of variables: $\q_{\mathcal{G}} = \sum_{i=1}^3 X_i$, $\q_m = X_1-X_2$, the flat spacetime singularity becomes manifest, and with the additional change, $\q_{\mathcal{G}} = \lambda_{\mathcal{G}} a_{\mathcal{G}}$ and $\q_m = \lambda$, one can again find the reparametrization of equation \eqref{eq:PF_operator} in the variables $(a_s,\,\lambda)$. 
Sending $\q_g\to 0$, the second order Picard-Fuchs operator factorizes in two linear differential operators:
\begin{align}
  &  \mathcal{L}_1=\frac{d}{d\lambda}+\frac{2 \lambda ^4+5 \lambda ^2-1}{(\lambda -1) \lambda  (\lambda +1) \left(\lambda ^2+1\right)} \ , \\
    & \mathcal{L}_1=\frac{d}{d\lambda}+\frac{\lambda }{(\lambda -1) (\lambda +1)} \, .
\end{align}
Consistently with the known answer of the one-loop three-site integral computation in flat space, 
as a non-trivial check, it is nice to verify that indeed, in the flat space limit, the elliptic subsector simplifies into polylogarithms.
In the study of this sector, algebraic letters appear of the form 
$(P + \sqrt{Q})/(P-\sqrt{Q})$,
with
\begin{align}
Q=& \,3 X_1^4+4 X_1^3 (X_2+X_3)-2 X_1^2 \left(7 X_2^2+2 X_2 X_3+3X_3^2\right) \nonumber \\
  & +4 X_1 (X_2-3 X_3) (X_2+X_3)^2\nonumber \\
  & +(3 X_2-5 X_3) (X_2+X_3)^3  \ . 
\end{align}

%% file: sections/outlooks.tex
\section{\label{sec:outlooks}Conclusion}

Loop corrections to cosmological observables are as challenging to compute as important for both theoretical and phenomenological reasons. In this paper we have begun a systematic study of the one-loop corrections to the Bunch-Davies wavefunction of the universe for the large class of scalar toy models which enjoys a first-principle definition in terms of cosmological polytopes. We exploit techniques coming from intersection theory and algebraic geometry, together with the structure of the underlying cosmological polytopes to set up the differential equations for the bubble and triangle integrals and extract information about the integrated functions. The one-loop correction to the two-site wave function falls in the realm of polylogarithmic functions, while, surprisingly, the one-loop correction to the three-site wave function involves also elliptic functions. The appearance of elliptic functions in the three-site graph is likely to extend to all polygon graphs with a higher number of sites. However, explicit computation needs to be carried out, to classify the space of functions appearing -- that might not be restricted to those emerging in the current analysis. In particular, it would be interesting to study whether there exists an underlying structure behind the differential equations for an arbitrary polygon, dictated by the geometry of both the integrand and the loop measure. This would allow predicting the space of functions at one-loop on general grounds. While our analysis for the bubble graph holds for arbitrary polynomial interactions in arbitrary FRW cosmologies, the one for the triangle graph holds in FRW cosmologies provided that the polynomial interactions are conformal. In these cases, just the integration over the loop edge weight needed to be considered. In order to extend it to more general interactions, the integration over the site weights ought to be considered as well. Also, note that the site weight integration for the two-site case was carried out directly. The differential equations for site weight integration by itself for loop graphs has been studied in \cite{He:2024olr} and turns out to return polylogarithms, in agreement with the expectations from the symbol analysis \cite{Arkani-Hamed:2017fdk, Hillman:2019wgh}. It would be interesting to consider both site and loop edge integrations on the same footing. 

The analysis can be further extended by considering different states running in the loop, such as both light and heavy scalars, as well as fermions. The latter has particular phenomenological value, as contributions from fermions arise at one-loop. Interestingly, the integrals appearing in the triangle graphs are of the same type as the ones studied in this paper \cite{Chowdhury:2024snc}. 

The vector space structure of cosmological integrals of the type we investigated in this paper, related to the wavefunction, also appear when spatial correlators are considered. For the latter case,  the loci given by the vanishing of the linear polynomial associated to the deletion of one edge of a loop graph, are absent \cite{AguiSalcedo:2023nds, Chowdhury:2023arc}. It would be interesting to understand how the absence of these singularities of the integrand affects the integrated result and, in particular, whether the evaluation of the latter also require elliptic functions, as well as how these features might be  dictated by the underlying combinatorics of the weighted cosmological polytopes \cite{Benincasa:2024lxe} -- these insights would sharpen our understanding on possible cancellation/simplifications emerging going from the wavefunction to the spatial correlations.

%% file: sections/appendix1.tex
\section{Linear relations via Intersection numbers}
\label{sec:A1}
The algebraic properties of period integrals of the type \eqref{eq:y_integral} can also be studied within intersection theory \cite{Mastrolia:2018uzb}, 
considering them as bilinear parings:
\begin{equation}
    \mathcal{I} = \int_\Gamma u \ \varphi  \equiv \langle \varphi \vert \Gamma ] , 
\end{equation}
between the differential $n$-form $\varphi$, viewed as an element of the $n$-th cohomology group for twisted co-cycles:
\begin{eqnarray}
    \langle \varphi \vert: \varphi \sim \varphi +\nabla_\omega\phi \ , 
\end{eqnarray}
where $\phi$ is a generic $(n-1)$ form and $\nabla_\omega = d + \omega \wedge$, with $\omega= d\log u$, and the integration domain $\vert \Gamma ]$, viewed as an element of the $n$-th homology group for twisted cycles.
The number of master integrals $\nu$ corresponds to the dimension of the co-homology groups, and can be obtained as explained in equation\eqref{eq:counting}.
In this case, choosing a basis of master integrals amounts to 
choose a basis of master forms $\{\langle e_i \vert\}_{i=1}^\nu$, 
such that ${\cal J}_i = \langle e_i | \Gamma]$, and the decomposition of equation \eqref{eq:ibpdec} can be seen as:
\begin{eqnarray}
    \langle \varphi \vert = \sum_{i=1}^\nu c_i \langle e_i \vert \ . 
    \label{eq:formdec}
\end{eqnarray}
Given a set of dual forms $\{\vert e_i^\vee \rangle\}_{i=1}^\nu$, belonging to the dual $n$-th co-homology group,
the coefficients $c_i$ of the decomposition of the integral \eqref{eq:formdec} can be obtained  by projection, dubbed {\it master decomposition formula}, as 
\begin{eqnarray}
    c_i = \sum_{j=1}^n\langle \varphi \vert e_j^\vee \rangle (C)^{-1}_{ji} \ , 
    \qquad C_{ij}= \langle e_i \vert e_j^\vee \rangle \ ,
\end{eqnarray}
by means of intersection numbers $\langle\cdot\vert \cdot \rangle$\cite{Mastrolia:2018uzb},
whose calculation can be performed using the methods developed in \cite{Frellesvig:2019uqt,Caron-Huot:2021xqj,Caron-Huot:2021iev,Fontana:2023amt,Brunello:2023rpq}. \\
Within this approach, the element of the matrix ruling the system of differential equations \eqref{eq:DEQ1} can be obtained from the master decomposition formula, in terms of intersection numbers, as:
\begin{eqnarray}
   ( {\mathbb A}_x)_{ij} = 
   \langle 
    u^{-1} \partial_x \, (u \, e_i)
    \, \vert \, 
   e^\vee_k\rangle \left( C^{-1}\right)_{kj} \ ,
\end{eqnarray}
where the resulting expression of ${\mathbb A}_x$ is the same as in \eqref{eq:DEQ1}.

\paragraph{One-loop two-site differential equations in terms of intersection numbers.}
The differential equations of Sections \ref{sec:Bubble} can also be derived using intersection numbers. After performing the change of variables $\{y_{12},y_{21}\}\to \{z_{\mathfrak{g}_1},z_{\mathcal{G}_{12}}\}:=\{\q_{\mathfrak{g}_1},\q_{\mathcal{G}_{12}}\}$, the polynomial $\bk$ appearing in the twist of the integral family defined in equation \eqref{eq:2site} reads:
\begin{align}
 \bk = &\left((z_{\mathfrak{g}_1}-\tilde{x}_1)^2-P^2\right) \left((z_{\mathfrak{g}_1}-z_{\mathcal{G}_{12}}+\tilde{x}_2)^2-P^2\right) \ . 
\end{align}
Choosing as variable ordering for the fibration: $\{z_{\mathfrak{g}_1},z_{\mathcal{G}_{12}}\}$, we get as dimensions for the inner and outer bases: 
\begin{equation}
    \nu^{(\mathcal{G}_{12})} = 2 , \qquad \nu^{(\mathfrak{g}_1\mathcal{G}_{12})}= 6 \ . 
\end{equation}
Choosing as bases:
\begin{align}
  &  e^{(\mathcal{G}_{12})} = \left\{1,\frac{1}{z_{\mathcal{G}_{12}}}\right\}, \nonumber \\  & e^{(\mathfrak{g}_{1}\mathcal{G}_{12})}= \left\{1,\frac{1}{z_{\mathfrak{g}_1}},\frac{1}{z_{\mathcal{G}_{12}}},\frac{z_{\mathfrak{g}_1}}{z_{\mathcal{G}_{12}}},\frac{1}{z_{\mathcal{G}_{12}}^2},\frac{1}{z_{\mathfrak{g}_1} z_{\mathcal{G}_{12}}}\right\}, 
\end{align}
the corresponding dual ones read:
\begin{align}
       & e^{\vee(\mathcal{G}_{12})} = \left\{1,\delta_{\mathcal{G}_{12}}\right\}, \nonumber \\
       & e^{\vee(\mathfrak{g}_{1}\mathcal{G}_{12})}= \left\{1,\delta_{\mathfrak{g}_{1}},\delta_{\mathcal{G}_{12}},z_{\mathfrak{g}_1} \delta_{\mathcal{G}_{12}},\omega_{\mathcal{G}_{12} }\delta_{\mathcal{G}_{12}},\delta_{\mathfrak{g}_{1}\mathcal{G}_{12}}\right\}, 
\end{align}
from which it is possible to obtain the system of differential equations. After rotating it to canonical form, one gets the system given in equation \eqref{eq:can_deq_1L}. \\ 

%% file: sections/appendix2.tex
\section{Details on the differential equations for the one-loop two-site graph}
\label{sec:A2}
In this appendix, we provide additional details on the integration of the one-loop two-site described in section \ref{sec:Bubble}. 
Starting from the original set of master integrals $\mathbfcal{I}=\{\mathcal{I}_{00},\mathcal{I}_{10},\mathcal{I}_{01},\mathcal{I}_{02},\mathcal{I}_{-11},\mathcal{I}_{11}\}$, it is possible to 
build a canonical basis $\mathbfcal{J} = \mathcal{R}.\mathbfcal{I}$, through the rotation matrix:
\begin{eqnarray}
\mathcal{R}=\left(
\scalemath{0.8}{
\begin{array}{cccccc}
 \frac{(2 \epsilon+1)^2}{P^2} & 0 & 0 & 0 & 0 & 0 \\
 0 & \frac{\epsilon (2 \epsilon+1)}{P} & 0 & 0 & 0 & 0 \\
 0 & 0 & \frac{\epsilon (4 \epsilon+1)}{P} & \frac{\epsilon (\tilde{x}_1+\tilde{x}_2)}{P} & 0 & 0 \\
 0 & 0 & 0 & -\epsilon & 0 & 0 \\
 -\frac{\epsilon (2 \epsilon+1)}{2 P (\tilde{x}_1+\tilde{x}_2)} & 0 & -\frac{\epsilon (2 \epsilon+1) (\tilde{x}_1-\tilde{x}_2)}{2 P (\tilde{x}_1+\tilde{x}_2)} & 0 &
   \frac{\epsilon (2 \epsilon+1)}{P (\tilde{x}_1+\tilde{x}_2)} & 0 \\
 0 & 0 & 0 & 0 & 0 & \epsilon^2 \\
\end{array}} 
\right)\nonumber  \\
\end{eqnarray}

The (constant) coefficient matrices appearing in equation \eqref{eq:can_deq_1L} read:
\begin{eqnarray}
&  & 
\mathbb{M}_1 = \scalemath{0.8}{\left(
\begin{array}{cccccc}
 4 & 0 & 0 & 0 & 0 & 0 \\
 0 & 2 & 0 & 0 & 0 & 0 \\
 0 & 0 & 2 & 0 & 0 & 0 \\
 0 & 0 & 0 & 0 & 0 & 0 \\
 0 & 0 & 0 & 0 & 2 & 0 \\
 0 & 0 & 0 & 0 & 0 & 0 \\
\end{array}
\right)} \ , \ \ 
\mathbb{M}_2 = \scalemath{0.8}{\left(
\begin{array}{cccccc}
 0 & 0 & 0 & 0 & 0 & 0 \\
 0 & 0 & 0 & 0 & 0 & 0 \\
 0 & 0 & 0 & 0 & 0 & 0 \\
 -\frac{1}{2} & 0 & 0 & 2 & 0 & 0 \\
 0 & 0 & 0 & 0 & 2 & 0 \\
 0 & 0 & 0 & 0 & 0 & 0 \\
\end{array}
\right)} \ , \nonumber \\ 
& & \nonumber \\
& &
    \mathbb{M}_3 = \scalemath{0.8}{\left(
\begin{array}{cccccc}
 0 & 0 & 0 & 0 & 0 & 0 \\
 \frac{1}{2} & 1 & 0 & 0 & 0 & 0 \\
 0 & 0 & 0 & 0 & 0 & 0 \\
 0 & 0 & 0 & 0 & 0 & 0 \\
 0 & 0 & 0 & 0 & 0 & 0 \\
 0 & 0 & \frac{1}{4} & \frac{1}{2} & \frac{1}{2} & 1 \\
\end{array}
\right)} \ , \ \ 
\mathbb{M}_4 = \scalemath{0.8}{\left(
\begin{array}{cccccc}
 0 & 0 & 0 & 0 & 0 & 0 \\
 0 & 0 & 0 & 0 & 0 & 0 \\
 0 & 0 & 0 & 0 & 0 & 0 \\
 0 & 0 & 0 & 0 & 0 & 0 \\
 0 & 0 & 0 & 0 & 0 & 0 \\
 0 & \frac{1}{2} & -\frac{1}{4} & -\frac{1}{2} & \frac{1}{2} & 1 \\
\end{array}
\right)} \ , \nonumber \\ 
& & \nonumber \\
& &
    \mathbb{M}_5 = \scalemath{0.8}{\left(
\begin{array}{cccccc}
 0 & 0 & 0 & 0 & 0 & 0 \\
 0 & 0 & 0 & 0 & 0 & 0 \\
 \frac{1}{2} & 0 & 1 & 2 & 0 & 0 \\
 \frac{1}{4} & 0 & \frac{1}{2} & 1 & 0 & 0 \\
 0 & 0 & 0 & 0 & 0 & 0 \\
 0 & 0 & 0 & 0 & 0 & 0 \\
\end{array}
\right)} \ , \ \ 
\mathbb{M}_6 = \scalemath{0.8}{\left(
\begin{array}{cccccc}
 0 & 0 & 0 & 0 & 0 & 0 \\
 -\frac{1}{2} & 1 & 0 & 0 & 0 & 0 \\
 0 & 0 & 0 & 0 & 0 & 0 \\
 0 & 0 & 0 & 0 & 0 & 0 \\
 0 & 0 & 0 & 0 & 0 & 0 \\
 0 & 0 & -\frac{1}{4} & \frac{1}{2} & -\frac{1}{2} & 1 \\
\end{array}
\right)} \ , \nonumber \\ 
& & \nonumber \\
& &
    \mathbb{M}_7 = \scalemath{0.8}{\left(
\begin{array}{cccccc}
 0 & 0 & 0 & 0 & 0 & 0 \\
 0 & 0 & 0 & 0 & 0 & 0 \\
 0 & 0 & 0 & 0 & 0 & 0 \\
 0 & 0 & 0 & 0 & 0 & 0 \\
 0 & 0 & 0 & 0 & 0 & 0 \\
 0 & -\frac{1}{2} & \frac{1}{4} & -\frac{1}{2} & -\frac{1}{2} & 1 \\
\end{array}
\right)} \ , \ \ 
\mathbb{M}_8 = \scalemath{0.8}{\left(
\begin{array}{cccccc}
 0 & 0 & 0 & 0 & 0 & 0 \\
 0 & 0 & 0 & 0 & 0 & 0 \\
 -\frac{1}{2} & 0 & 1 & -2 & 0 & 0 \\
 \frac{1}{4} & 0 & -\frac{1}{2} & 1 & 0 & 0 \\
 0 & 0 & 0 & 0 & 0 & 0 \\
 0 & 0 & 0 & 0 & 0 & 0 \\
\end{array}
\right)} \ , \nonumber \\ 
\end{eqnarray}
The boundary conditions can be fixed as follows: 
\begin{itemize}
\item $\mathcal{J}_1$,$\mathcal{J}_5$ can be obtained by direct integration, since in momentum space they appear as massless two-points functions $\mathcal{G}_{a_1,a_2}$ with half-integer exponents in Euclidean space, of the type of equation \eqref{eq:flat_bubble}
     whose solution is known \cite{Foffa:2016rgu}.
\item $\mathcal{J}_2$,$\mathcal{J}_4$,$\mathcal{J}_6$ can be fixed order by order in $\epsilon$ imposing regularity conditions in the spurious poles, at the point $(\tilde{x}_1,\tilde{x}_2,P)=(1,1,1)$.
\item $\mathcal{J}_4$ can be fixed by matching with the direct integration at $(\tilde{x}_1,\tilde{x}_2,P)=(0,0,1)$, which at this point becomes of the form of integral of equation \eqref{eq:flat_bubble}.
\end{itemize}
\vspace{1in}